\documentclass[a4paper,11pt]{article}
\pdfoutput=1
\usepackage{jcappub}

\usepackage{amsmath}
\usepackage{amssymb}
\usepackage{amsfonts}
\usepackage{mathtools}

\usepackage{lineno}
\usepackage{lscape}
\usepackage{soul}
\usepackage{subcaption}
\usepackage{graphicx}
\usepackage{float}
\usepackage{color}
\usepackage{bm}
\usepackage{epstopdf}
\usepackage{url}
\usepackage{booktabs}
\usepackage{mdframed}
\usepackage{hyperref}
\usepackage{booktabs}
\usepackage{siunitx}
\usepackage{multirow}
\usepackage{appendix}

\usepackage{changes}

\usepackage[utf8x]{inputenc}
\usepackage[normalem]{ulem}

\definecolor{burgundy}{rgb}{0.6, 0.0, 0.0}
\definecolor{persianblue}{rgb}{0.11, 0.22, 0.73}
\hypersetup{colorlinks, citecolor=burgundy, linkcolor=persianblue, urlcolor=persianblue}

\definecolor{Gray}{gray}{0.95}

\title{Features in cosmic-ray lepton data unveil the properties of nearby cosmic accelerators}

\author[a,b,d,1]{O. Fornieri,\note{Corresponding author.}}
\author[d]{D. Gaggero}
\author[b,c]{and D. Grasso}

\affiliation[a]{Dipartimento di Scienze Fisiche, della Terra e dell'Ambiente, Universit\`a di Siena,\\Via Roma 56, 53100 Siena, Italy}
\affiliation[b]{INFN Sezione di Pisa,\\Polo Fibonacci, Largo B. Pontecorvo 3, 56127 Pisa, Italy}
\affiliation[c]{Dipartimento di Fisica, Universit\`a di Pisa,\\Polo Fibonacci, Largo B. Pontecorvo 3}
\affiliation[d]{Instituto de F\'{i}sica Te\'{o}rica UAM-CSIC,\\Campus de Cantoblanco, E-28049 Madrid, Spain}

\emailAdd{ottavio.fornieri@pi.infn.it}
\emailAdd{daniele.gaggero@uam.es}
\emailAdd{dario.grasso@pi.infn.it}

\abstract{We present a comprehensive discussion about the origin of the features in the leptonic component of the cosmic-ray spectrum. 
Working in the framework of a up-to-date CR transport scenario tuned on the most recent AMS-02 and Voyager data, we show that the prominent features recently found in the positron and in the all-electron spectra by several experiments are compatible with a scenario in which pulsar wind nebulae (PWNe) are the dominant sources of the positron flux, and nearby supernova remnants (SNRs) shape the high-energy peak of the electron spectrum.  
In particular we argue that the drop-off in positron spectrum found by AMS-02 at $\sim 300$ GeV can be explained --- under different assumptions --- in terms of a prominent PWN that provides the bulk of the observed positrons in the $\sim 100$ GeV domain, on top of the contribution from a large number of older objects. Finally, we turn our attention to the spectral softening at $\sim 1$ TeV in the all-lepton spectrum, recently reported by several experiments, showing that it requires the presence of a nearby supernova remnant at its final stage.}

\begin{document}
\maketitle
\flushbottom

\section{Introduction}\label{sec:intro}

The origin and transport properties of leptonic cosmic rays (CRs) have intrigued scientists for decades.
Differently from the nuclear components of CRs, the propagation of CR leptons in the interstellar medium (ISM) of the Galaxy is characterized not only by a diffusive propagation but also by strong energy losses which --- above few GeV --- are dominated by synchrotron emission and Inverse Compton (IC) scattering. 

Those losses, that are at work during both the acceleration process in the source environment and the propagation in the ISM, are likely responsible for the steeper spectrum, observed for CR electrons, with respect to that of CR hadrons, as well as for its lower maximal energy.
Furthermore, the presence of a transition from a diffusion-dominated to a loss-dominated regime in the ISM is expected to produce a cooling break in the propagated spectrum. 

The accurate measurement of such features in the CR lepton spectrum --- which may be observed either directly at Earth or indirectly by looking for its imprint on the synchrotron and IC diffuse emission --- may offer valuable clues on the ages/positions of the sources, as well as on the details of CR transport. 
Also, given the $\propto E^2$ scaling of the leptonic energy-loss rate, the effective {\it horizon} associated to CR leptons progressively shrinks as energy increases, hence the stochastic nature of the sources is expected to play a more and more important role with increasing energy. This trend implies even more pronounced features at high energies, as noticed already in \cite{Shen1970ApJ...162L.181S} and further elaborated in more recent times \cite{1995A&A...294L..41A,2004ApJ...601..340K}.

On the experimental side, several experiments have recently provided accurate measurements of the leptonic ($e^{-}, e^{+}$ and $e^{+} + e^{-}$) spectra up to $\sim \mathcal{O}(10)$ TeV and have revealed significant features. It is therefore challenging to connect these observed features to the physics arguments discussed above.

Regarding the electrons, we remark in particular that the AMS-02 spectrum exhibits a hardening at $\simeq 40$ GeV \cite{AMS2019PhRvL.122j1101A}. At even higher energies, H.E.S.S. \cite{2009A&A...508..561A,kerszberg_ICRC}, DAMPE \cite{Ambrosi:2017wek} and CALET \cite{Adriani:2018ktz} measured the $e^- + e^+$ spectrum up to $\simeq 20$ TeV and outlined a sharp softening at $\simeq 1$ TeV.
Above that energy, the power-law spectrum extends, with no clear sign of cutoff, all the way up to the maximal detected energy. 
Given the limited distance electrons can propagate in the multi-TeV energy range, the interpretation of these results is very challenging and triggered a debate in the scientific community. 
A non-conventional scenario, where the diffusion coefficient is decoupled by secondary/primary ratios, has been proposed \cite{Lipari:2018usj} where the TeV spectral feature was interpreted as a {\it cooling break}, associated to a much shorter CR residence time in the Galaxy compared to the conventional estimates. However, this interpretation seems to have problems explaining the low-energy flux consistently.

Another valuable piece of information comes from the CR positron flux. A guaranteed flux of $e^+$ is expected due to the interaction of CR nuclei (mainly protons and Helium) with the ISM gas. This component is expected to decrease with respect to the $e^- + e^+$ flux as energy increases. However, the discovery of an increasing positron fraction above 10 GeV by PAMELA \cite{2009Natur.458..607A}, later confirmed and better characterized by AMS-02 \cite{Aguilar:2013qda}, was then corroborated by the measurement of the absolute $e^+$ spectrum by both experiments \cite{2013PhRvL.111h1102A,Aguilar:2014mma} showing that the anomaly cannot be attributed to a steeper-than-expected $e^-$ spectrum and that a primary origin of Galactic high-energy positrons needs to be identified. As discussed in a long series of papers (see \textit{e.g.} \cite{HardingRamaty1987ICRC,Hooper:2008kg,2009APh....32..140G,2011ASSP...21..624B}), the electron+positron pairs accelerated at pulsar wind nebulae (PWNe) provide a reasonable explanation for this flux, both in terms of energy budget and spectral shape.
We notice that a scenario invoking PWNe as the origin of the positron excess has recently been further debated after the detection of TeV gamma-ray halos around the Geminga and Monogem nearby pulsars by HAWC \cite{2017Sci...358..911A} and by Fermi-LAT \cite{2019arXiv190305647D}, which has been interpreted in terms of IC emission from a fresh population of electrons and, plausibly, positrons \cite{2017PhRvD..96j3013H}, confined in the vicinity of those pulsars. Also, recent studies conducted on bow-shock wind nebulae associated to a nearby ($\sim 150$ pc) millisecond pulsar (BSWN) discuss the contribution to the positron excess coming from those compact objects, and to the all-lepton flux as well coming from the shocked medium \cite{Bykov_2019}. Finally, several times in the literature, outflows of relativistic leptons have been reported in correspondence to fast neutron stars: see for instance the Guitar Nebula \cite{CordesNature} and the Lighthouse nebula \cite{Pavan:2013lka}.

However, a unified picture of all the observable leptonic and hadronic channels built on these ideas, and based on an up-to-date transport scenario, is still lacking.

In this paper we propose a significant step forward towards such a picture and provide a comprehensive, state-of-the-art discussion about the origin of these spectral features and their connection with the physical properties of the nearby accelerators. 

The key elements of novelty of our work are the following:

\begin{itemize}
\item Regarding the interpretation of the positron flux, in the context of the PWN-origin scenario
we account for the large and often unaccounted uncertainties due to the unknown details of the emission process
(unknown acceleration spectrum; unknown duration of the emission).
For this reason, we perform different Bayesian fits to constrain the injection parameters under very different assumptions on the injection history (burst-like or continuous) and on the spectral shape, in order to bracket these extreme uncertainties.   
\item Regarding the interpretation of the all-lepton flux, given the wider context of an updated transport scenario tuned on all the available data sets, and given the aforementioned PWN contribution, we implement some realistic realizations of the scenario
proposed in \cite{Recchia:2018jun} in which the $e^\pm$ flux above the TeV is dominated by the emission of a hidden, middle-aged remnant with declining luminosity.
After a careful assessment of the contribution of the known nearby supernova remnants, we show that the emission of such hidden SNR is required to reproduce the spectral feature reported by H.E.S.S. and characterize its properties.
\end{itemize}

The structure of the paper is the following. 
We first identify a transport scenario that provides a satisfactory description of light-nuclei CR-data released by AMS-02 mostly, which is required to fix the diffusion parameters that will enter in determining the shape and features of the propagated lepton spectra. 
Then, we turn our attention to the positron flux and model its observed spectrum in terms of i) a conventional secondary component produced by hadronic spallation, ii) a primary extra-component that dominates at intermediate energies and originates by a large number of distant, old pulsars, iii) and one or few nearby pulsars as the main possible contributors at high-energies.
Finally, we concentrate on the electron and all-lepton data and analyze the contributions from nearby asymmetric accelerators within the same transport scenario.

\section{Characterization of the large-scale CR transport scenario}\label{sec:characterizazione_CR_transport}

In this section we settle the CR transport setup that will be adopted throughout the paper. It is understood that this setup captures an effective large-scale average of transport properties and is not affected by strong local fluctuations, as those identified for instance in \cite{2017Sci...358..911A}. 

\subsection{Description of the setup}\label{subsec:Spatially_continuous_and_stationary_sources}

The usual starting point is the phenomenological equation that captures CR diffusion in space and momentum, energy losses, advection, reacceleration, nuclear spallation and decay \cite{Ginzburg1964,Ginzburg:1990sk}:

\begin{equation}\label{eq:prop}
\begin{split}
- {\vec{\nabla}} \cdot ( D \, \vec{\nabla} N_i \,+\, \vec{v}_w N_i) + \frac{\partial}{\partial p} \left[ p^2 D_{pp} \, \frac{\partial}{\partial p} \left( \frac{N_i}{p^2} \right) \right] - \frac{\partial}{\partial p} \left[ \dot{p} N_i - \frac{p}{3} \left(\vec{\nabla} \cdot \vec{v}_w \right) N_i \right] = \\
Q + \sum_{i<j} \left( c \beta n_{\rm gas} \, \sigma_{j \rightarrow i} + \frac{1}{\gamma \tau_{j \rightarrow i}} \right) N_j - \left( c \beta n_{\rm gas} \, \sigma_i + \frac{1}{\gamma \tau_i} \right) N_i
\end{split}
\end{equation}
for which we remind to \cite{2017JCAP...02..015E} to have a detailed description of each term.

The equation above is solved for all the relevant species with the publicly available version of {\tt DRAGON}\footnote{\url{https://github.com/cosmicrays/DRAGON}} \cite{Evoli:2008dv,2017JCAP...02..015E}, and the following settings are specified:

\begin{itemize}
\item For CR nuclei and protons, we adopt a continuous source distribution in two dimensions --- cylindrical symmetry --- taken from \cite{Ferriere:2001rg}. Such distribution accounts for the spatial distribution of both type Ia (traced by old stars in the disk), and type II (traced by pulsars) supernovae.

The injection spectrum of each species is modeled as a broken power law in rigidity. Two breaks are introduced to effectively reproduce low-energy data as well as the hardening measured by AMS-02 and other experiments at few hundred GeV/n. This will be discussed in more detail in Section \ref{subsec:setting_CR_nuclei_parameters}.

\item For the aim of this study we assume uniform and isotropic CR diffusion characterized by a spatially-independent scalar diffusion coefficient that exhibits the following scaling as a function of rigidity:

\begin{equation}\label{eq:diff_coeff}
    D_{xx}(\rho) = D_0 \left( \frac{\rho}{\rho_0} \right)^{\delta} 
\end{equation}
where $D_0 \equiv D_{xx}(\rho_0)$ is the diffusion coefficient normalization at a reference rigidity, that in this work we fix at $\rho_0 = 1$ GV.

\item We account for diffusive reacceleration adopting a finite value of the Alfv{\`e}n velocity $v_{\textrm{A}}$ which fixes the diffusion coefficient in momentum space through the expression \cite{Ginzburg:1990sk,Drury:2015zma}

\begin{equation}\label{eq:diff_coeff_mom}
    D_{pp}(p) = \frac{v_{\textrm{A}}^2~ p^2}{\delta(4 - \delta)(4 - \delta^2) D_{xx}(p)} \, .
\end{equation}

\item Hadronic energy losses (ionization and Coulomb scattering) as well as Bremsstrahlung, synchrotron and Inverse Compton for leptons are considered, as implemented in the public {\tt DRAGON} code~\cite{2017JCAP...02..015E}.

The astrophysical ingredients relevant for the computation of the energy-loss term are the following:

\begin{itemize}

\item A smooth, cylindrically symmetric gas distribution, taken from \cite{1976ApJ...208..346G,Bronfman1988} and implemented in the standard public versions of both {\tt GALPROP} \cite{Galprop2,Galprop3} and {\tt DRAGON}.

\item Magnetic field as parametrized in \cite{Pshirkov:2011um}; accordingly, we set $B_{\textrm{0,disk}} = 2 \, \mu \text{G}$, and $B_{\textrm{0,Halo}} = 4 \, \mu \text{G}$. For the turbulent field, we adopted the parametrization described in \cite{2017JCAP...02..015E} (Equation C.6) with $B_{\textrm{0,turb}} = 7.5 \, \mu \text{G}$.

\end{itemize}

\item Concerning the spallation ($\sigma_{i \rightarrow j}$) and inelastic-scattering ($\sigma_i$) cross-sections, in order to allow a more direct comparison with most of the related literature, we use the routines implemented in the standard public version of {\tt GALPROP}.

\end{itemize}

\subsection{Setting source and transport parameters against CR nuclei data}\label{subsec:setting_CR_nuclei_parameters}

While gas density, magnetic- and interstellar-radiation-field distributions are fixed (though with some uncertainties) on the basis of astronomical data, CR injection spectra and diffusion parameters are largely unknown and have to be settled by comparing {\tt DRAGON} predictions with CR data. 
We use here AMS-02 data for almost all species~\cite{PhysRevLett.114.171103,PhysRevLett.119.251101} and for the B/C ratio~\cite{PhysRevLett.117.231102} complemented with Voyager data~\cite{Cummings_2016} for low-energy protons and other nuclei. Finally, HEAO-3~\cite{1989ApJ...346..997B} data are considered to determine the normalization of nuclear species heavier than Nitrogen.

Voyager data (below 1 ${\rm GeV/n}$) are collected outside the Heliosphere, allowing us to tune the low-energy injection spectra without being affected by solar modulation (parametrized in terms of a modulation potential $\phi_{\textrm{mod}}$).

Once the injection spectra are fixed, we are able to constrain the value of $\phi_{\textrm{mod}}$ by fitting the low-energy ($\lesssim 10$ GeV/n) AMS-02 modulated data. The values of $\phi_{\textrm{mod}}$ that we obtain are consistent with the independent measurement performed at ground-based detectors~\cite{2005JGRA..11012108U,2011JGRA..116.2104U}.

With this cross-checked estimation of $\phi_{\textrm{mod}}$ at hand, we can connect with the intermediate-energy ($>10$ GeV/n) AMS-02 points, and conclude that a first break at low energy ($\lesssim 10$ GeV/n) is required to reproduce the proton/nuclei data. 
This procedure is very important because {\bf a)} it justifies the presence of a low-energy break also in the $e^{-}$ spectrum (although in this paper we are agnostic about its physical origin) if we consider a common origin for CR protons/nuclei and electrons (\textit{e.g.} SNRs); {\bf b)} it validates the values used for the modulation potential, which significantly affects the leptonic spectra all the way up to $\sim 30$ GeV. 

A second break has to be implemented in the hadronic species at a few hundred of GeV, as reported by the AMS-02 observations cited above. 
The origin of this break is still under debate. However, the more pronounced hardening found in secondary nuclei seems to point towards a diffusive origin, and the physical interpretations proposed so far deal with a different nature \cite{Evoli:2018nmb} (or behaviour~\cite{Yan:2002qm,Yan:2004aq,Yan:2007uc}) of the turbulent cascade in the halo and in the disk. 
Given that the propagated spectrum of the primaries scales as $\sim E^{- \Gamma_{\textrm{inj}} - \delta}$ above $\mathcal{O}(10)$ GeV, we effectively mimic the diffusive break with a break in the injection. 

On the other hand, we emphasize that a completely different treatment is required for primary leptons: in fact, at $\mathcal{O}(100)$ GeV, leptons are mostly coming from the local region, so they spend most of their time in the same galactic environment. For this reason, we choose to model the smooth leptonic component as a single power-law in rigidity above $\sim 10$ GeV, as it will be seen in Section \ref{sec:SNR_electrons}.

\begin{figure}
    \centering
    \begin{subfigure}{.495\linewidth}
        \centering
        \includegraphics[width=1.\linewidth]{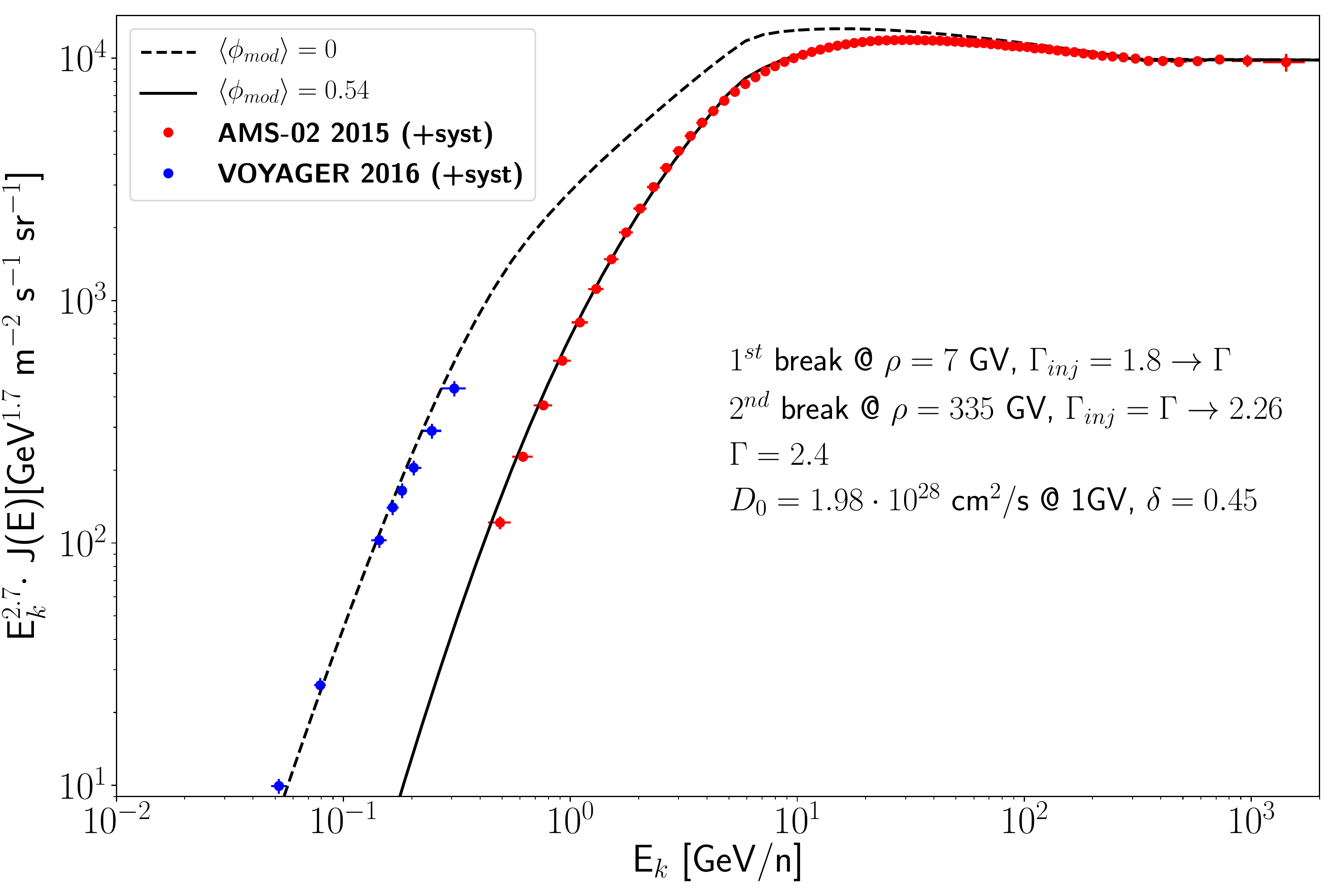}
        \caption{}
        \label{subfig:protons}
    \end{subfigure}
    \begin{subfigure}{.495\linewidth}
        \centering
        \includegraphics[width=1.\linewidth]{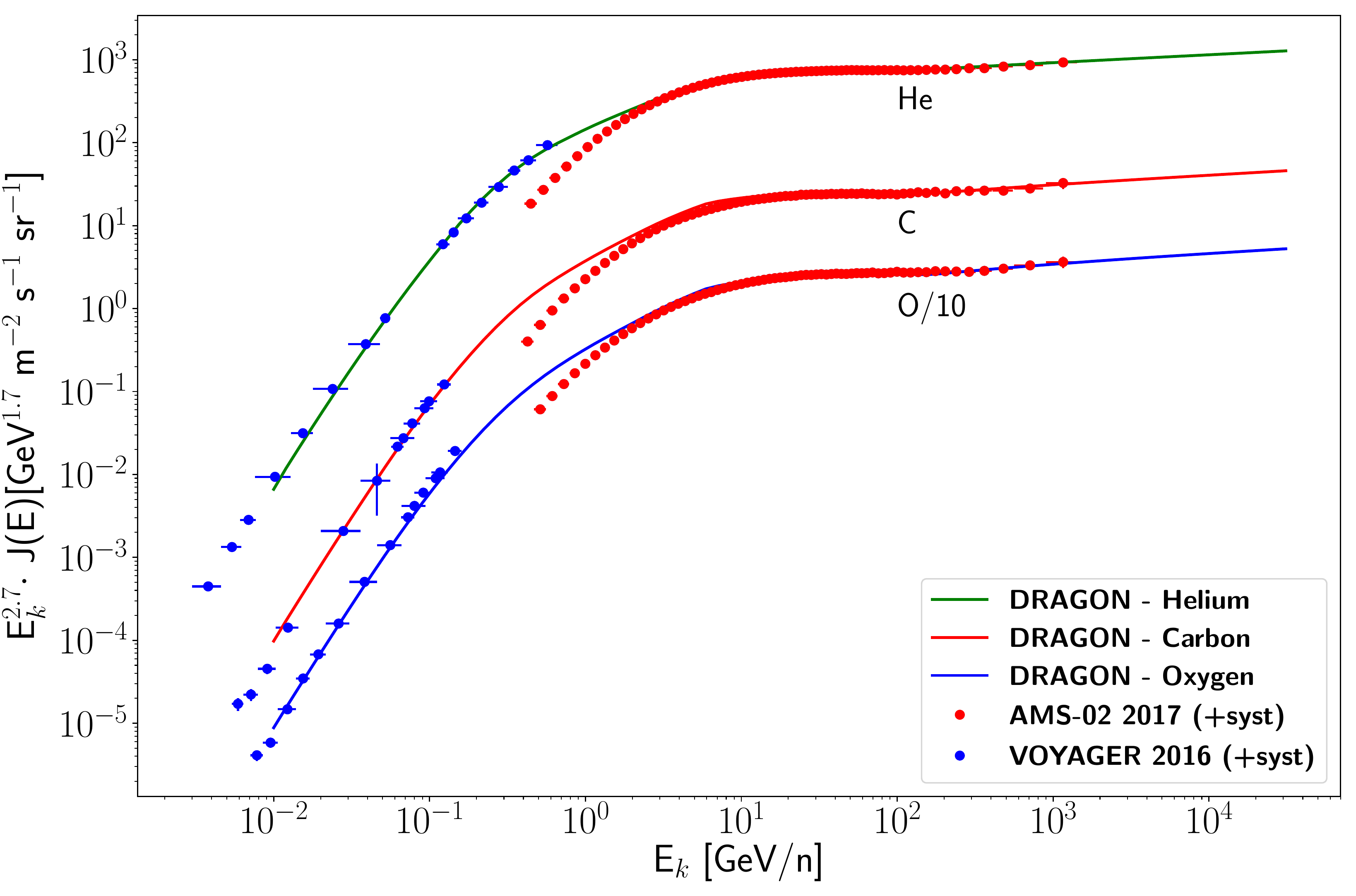}
        \caption{}
        \label{subfig:He_C_O_nuclei}
    \end{subfigure}
    \begin{subfigure}{.495\linewidth}
        \centering
        \includegraphics[width=1.\linewidth]{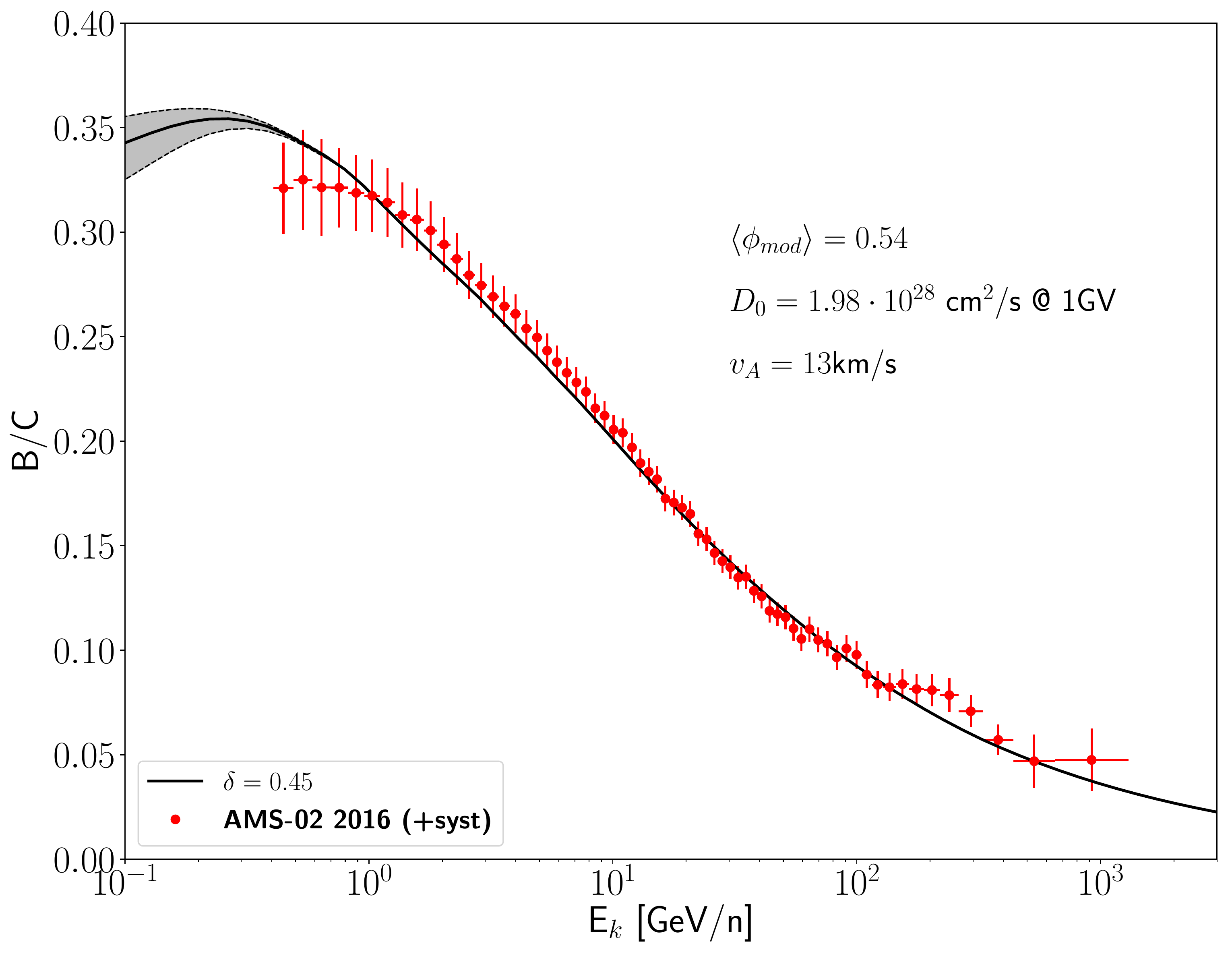}
        \caption{}
        \label{subfig:B-C_different_diffusion_deltas_vA13}
    \end{subfigure}
    \begin{subfigure}{.495\linewidth}\label{subfi}
        \centering
        \includegraphics[width=1.\linewidth]{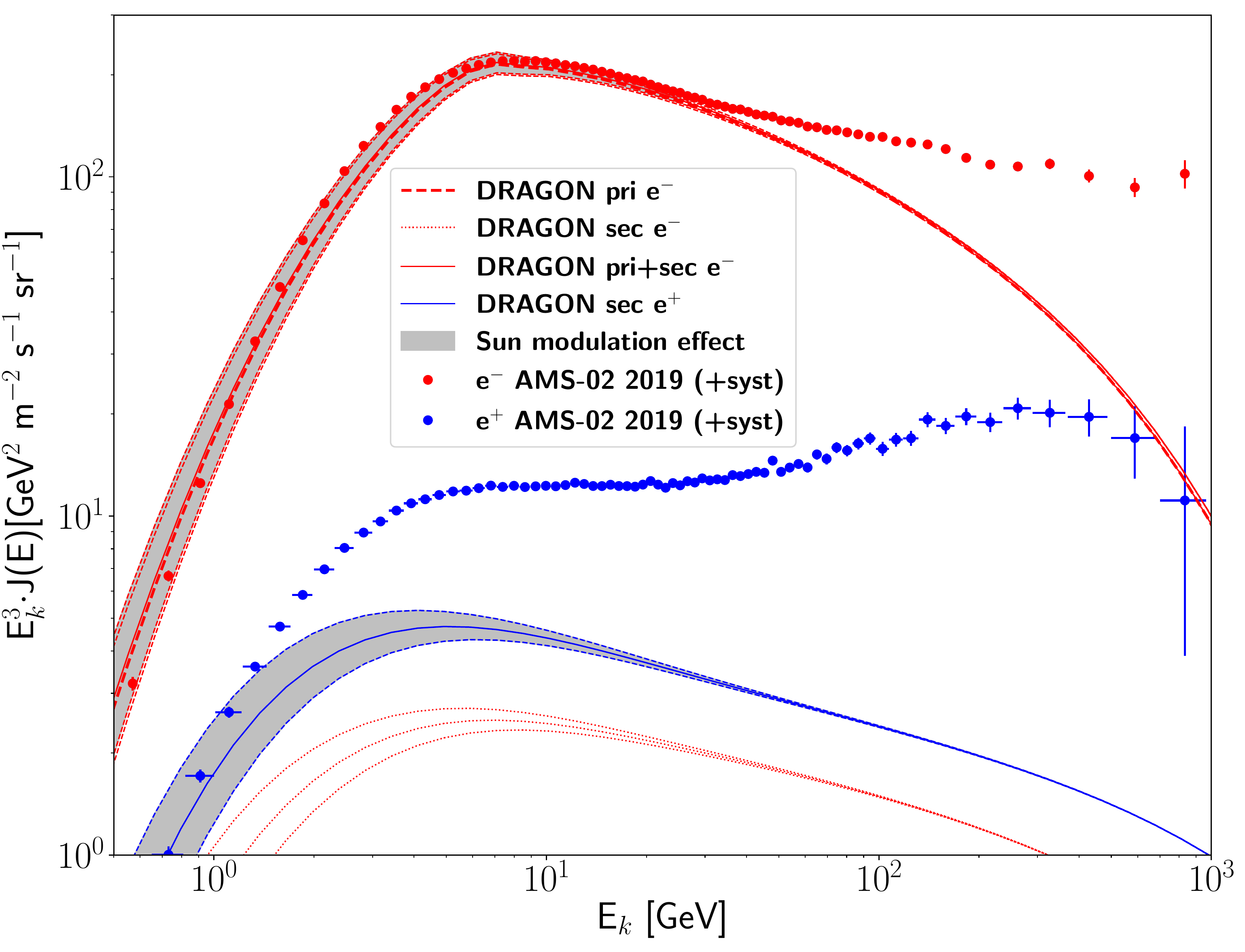}
        \caption{}
        \label{subfig:electrons_and_positrons_AMS}
    \end{subfigure}
    \caption{\small{The propagated spectra computed with our reference model of (a) protons, (b) Helium, Carbon and Oxygen (Oxygen flux is divided by 10 for clarity) are compared with AMS-02 \cite{PhysRevLett.114.171103,PhysRevLett.119.251101} (accounting for solar modulation) and Voyager \cite{Cummings_2016} (interstellar) data. For Voyager C and O data, data points may overlap due to measurements coming from different telescopes and modes (denoted as TT in the reference). In (c) the B/C ratio is computed for the same model and is plotted against AMS-02 experimental data \cite{PhysRevLett.117.231102}. (d) Primary and secondary $e^{-}$ and $e^{+}$ spectra computed with {\tt DRAGON} accounting only for the contribution of distant SNRs and secondary production in the ISM. The red and blue dots are AMS-02 experimental data \cite{AMS2019PhRvL.122j1101A,Aguilar:2019owu}. The silver band accounts for the solar modulation $\langle \phi_{\textrm{mod}} \rangle = 0.54 \pm 0.10$, estimated according to \cite{2005JGRA..11012108U,2011JGRA..116.2104U} for the whole period of data taking.}}
    \label{fig:background_plots}
\end{figure}

In order to implement what discussed above, we performed several two-dimensional {\tt DRAGON} runs in a grid with 41 linearly spaced points along the radial axis $R \in [0, 12] \, \textrm{kpc}$ and 81 linearly spaced points in the vertical axis $z \in [0, \pm 4] \, \textrm{kpc}$, where we propagated particles of energy $E_{\textrm{k}} \in [10 \, \textrm{MeV}, 30 \, \textrm{TeV}]$, logarithmically spaced according to $E_{\textrm{k}}[i] = \exp( \ln{(E_{\textrm{k,min}})} + i \, \ln{(E_{\textrm{k,factor}})})$, where $E_{\textrm{k,factor}} = 1.2 \, \textrm{GeV}$. Based on this setup, we identify a satisfactory scenario, characterized by the parameters listed in Table \ref{tab:hadronic_best_fit}. As reported there, and also shown in Figures \ref{subfig:protons} and \ref{subfig:He_C_O_nuclei}, the observed spectra are reproduced introducing a low-energy break at 7 ${\rm GeV/n}$, for all species, and a high-energy hardening at 335(165) ${\rm GeV/n}$ for protons (heavier nuclei).
\begin{table}[t]
    \centering
    \begin{tabular}{|c|c|c|c|c|c|c|c|c|}
    \hline
         & $\bm{v_{\textrm{A}}}$ [km/s] & $\bm{D_{0}}$ [$\text{cm}^2$/s] & $\bm{\delta}$ & $\bm{\Gamma_{\textrm{inj,l}}}$ & $\bm{E_{\textrm{b,1}}}$ [GeV/n] & $\bm{\Gamma_{\textrm{inj,m}}}$ & $\bm{E_{\textrm{b,2}}}$ [GeV/n] & $\bm{\Gamma_{\textrm{inj,h}}}$ \\
         \hline 
         \hline
         \textbf{p} & \multirow{4}{*}{13} & \multirow{4}{*}{$1.98 \cdot 10^{28}$} & \multirow{4}{*}{0.45} & 1.8 & \multirow{4}{*}{7} & 2.4 & 335 & 2.26 \\ 
         \cline{5-5} \cline{7-9}
         \textbf{He} &  &  & & 2.0 & & 2.28 & 165 & 2.15 \\  
         \cline{5-5} \cline{7-9}
         \textbf{C} &  &  & & 2.0 & & 2.38 & 165 & 2.15 \\
         \cline{5-5} \cline{7-9}
         \textbf{O} &  &  & & 2.0 & & 2.38 & 165 & 2.15 \\
         \hline
    \end{tabular}
    \caption{\small{The table reports the parameters of the source distribution and of the reference transport model. The labels (l,m,h) refer to \textit{low}, \textit{medium} and \textit{high} energy injection indices.}}
    \label{tab:hadronic_best_fit}
\end{table}

It should be noted that an approximate degeneracy holds between the diffusion coefficient normalization and the diffusive-halo height-scale $H$ since the CR escape time, hence the secondary/primary ratio, only depends on the ratio $D_0/H$. In this paper we use $H = 4$ kpc.
We notice that a different choice of $H$ within a wide range of allowed values has no significant effect on the electron spectrum and may affect the positron spectrum only below $\sim 10$ GeV (see Figure 4 in \cite{2013JCAP...03..036D}) with no impact on the results of this analysis.

Similarly to the results reported in \cite{DiBernardo:2009ku}, and --- more recently --- in \cite{Yuan:2017ozr,Genolini:2019ewc}, the B/C ratio is nicely matched for a value of $\delta$ close to $0.45$. Performing a statistical analysis aimed at the determination of the uncertainties in the propagation parameters, involving the full set of secondary/primary ratios, is beyond the aims of this work.
We mention however that varying the main parameters in the small allowed ranges found in \cite{Genolini:2019ewc} would have no relevant impact on the electron and positron spectra and therefore on the conclusions of this paper.

\subsection{Primary electrons and secondary positrons}\label{sec:SNR_electrons}

In the standard CR transport scenario, the Galactic SNRs are expected to generate the bulk of the observed CR electrons as well. Moreover, a guaranteed source of secondary electrons and positrons is provided by the scattering of CR nuclei --- mostly protons and $^4{\rm He}$ --- with the ISM gas.

For what concerns the primary electrons, we remark that, although the acceleration mechanism is expected to be the same as the one at work for the nuclear species, the injection spectrum into the ISM should be steeper (with $\Delta \Gamma$ as large as $\sim 0.4$) due to synchrotron losses in the SNR magnetic field, which is also amplified by CR-induced turbulence \cite{Diesing:2019lwm}. 
We notice that the {\tt DRAGON} output is in good agreement with analytical computations \cite{1974Ap&SS..29..305B,Lipari:2018usj} predicting a propagated spectral index $\displaystyle{\Gamma = \Gamma_{{\rm inj}} + \frac{\delta}{2} + \frac{1}{2}}$ above few GeV. We compute the propagated spectra at Earth with {\tt DRAGON} adopting the setup derived in the previous paragraph and implementing an electron injection spectrum $\Gamma^e_{\rm inj} = 2.7$ ($1.6$) above (below) $7$ GeV. This allows to reproduce the measured spectrum up to $\sim 50$ GeV, above which it displays a pronounced hardening (see Figure \ref{subfig:electrons_and_positrons_AMS}).
In our opinion that feature corresponds to the expected breakdown of the assumption of a continuous, steady-state source term that characterizes the large-scale models developed with {\tt DRAGON.}
Indeed, the mean distance of active SNRs from the Earth is expected to be of few kpc's. As a consequence, we expect that already above $\sim 100$ GeV the energy losses will limit the number of SNRs contributing to the observed CR electron flux to just a few. The contribution of individual CR electron sources will be discussed in detail in Section \ref{sec:local_SNR}.

Regarding the secondary positrons, they are computed with {\tt DRAGON} as well, within the same transport setup. The result is also reported in Figure \ref{subfig:electrons_and_positrons_AMS}. The plot clearly shows evidence of the well known \textit{positron excess} above $\sim 40$ GeV, pointed out since the first release of the PAMELA data \cite{Adriani:2008zr}. However, differently from other previous works (see \textit{e.g.} \cite{Galprop2,Galprop3}), we find an excess at all energies above $\sim 1$ GeV: this is consistent with other dedicated analyses, such as \cite{Boudaud:2016jvj}.

Even though alternative CR propagation scenarios may be invoked to account for the unexpected production of positrons \cite{2017PhRvD..95f3009L}, as well as interpretations based on dark matter annihilation (see for instance the recent review \cite{Gaggero:2018zbd} and references therein), lepton pair emission from pulsar wind nebulae seems to be a more natural candidate.
We will assess their contribution in the next paragraph.

\section{The positron excess}

In this section we focus on positron data and present a detailed discussion on their possible interpretation. In particular we address from the phenomenological point of view the role of local and distant sources of relativistic electron+positron pairs, such as pulsar wind nebulae (PWNe): we discuss whether a scenario in which the positron flux is dominated by this class of sources is viable (both from the point of view of the energy budget and of the spectral features) and assess whether the current data allow us to pinpoint which PWNe are most likely to contribute in the different energy ranges.

\subsection{Setting the stage: basic aspects of pulsar acceleration in pulsar wind nebulae and relevant caveats}\label{subsec:Setting_the_stage}

Pulsar wind nebulae are structures born inside the shells of supernova remnants, which emit a broad-band spectrum of non-thermal radiation powered by fast-spinning magnetized neutron stars with a typical radius $R \sim 10$ km and periods of $\mathcal{O}(0.1 \,-\, 10)\,{\rm s}$, typically detected in the radio and/or gamma-ray band as {\it pulsars}.

As mentioned in the Introduction, the role of pulsars and PWNe as relevant and efficient antimatter factories in the form of $e^{\pm}$ pairs and their contribution to the detected all-lepton flux have been debated for a long time in the literature, since the pioneering works of the past century \cite{Shen1970ApJ...162L.181S,HardingRamaty1987ICRC,1995PhRvD..52.3265A}. 
We will recall in this section some important aspects of the physics that characterizes these objects, in order to motivate our phenomenological parameterization of the problem.

To characterize the emission from a PWN, it is important to assess: 1) the energy release as a function of time, and 2) the acceleration mechanisms of the electron+positron pairs, hence the energy spectrum of such leptons when they are eventually released in the interstellar medium (ISM).

\begin{enumerate}

\item Regarding the former, we recall that the pulsar spin-down is usually described by the following model-independent equation:
\begin{equation}\label{eq:energy_loss_pulsar}
    \dot{\Omega}(t) = - \kappa_{0} \cdot \Omega(t)^{n},
\end{equation}
where $\Omega(t) = P^{-1}(t)$ is the rotation frequency, $\kappa_{0}$ and $n$ are parameters that depend on the specific energy-loss process; in particular $n$ is commonly called \textit{braking index}.

This equation can be solved to get $\Omega(t)$ and the time evolution of the luminosity, which, in terms of the conversion efficiency ($\eta^{\pm}$) of the released energy into $e^{\pm}$ pairs, can be written as follows:
\begin{equation}\label{eq:decaying_luminosity}
L(t) = I \Omega(t) \dot{\Omega}(t) \,=\, \frac{\eta^{\pm} L_{0,\gamma}}{ \left(  1 + \frac{t}{\tau_{0}}  \right)^{\frac{n+1}{n-1}}}
\end{equation}
where $\displaystyle{ \tau_0 \,\equiv\, \frac{1}{(n-1) \kappa_0 \Omega_0^{n-1}}} \,$ and $t$ is the age of the source.

Under the assumption that at present time the pulsar rotation period is $P(t) \gg P_0 \equiv P(t = 0)$, we can approximate $t$ with its {\it characteristic age}, $\displaystyle{t_{\textrm{ch}} \approx \frac{P}{(n-1)\dot{P}}}$ \cite{2005handbook}.

According to \eqref{eq:decaying_luminosity}, the release process is regulated by the ratio $t_{\textrm{ch}}/\tau_0$. When $t_{\textrm{ch}}/\tau_0 \ll 1$, we can Taylor-expand the function $L(t) \approx \eta^{\pm} L_{0,\gamma} \, (1 - \frac{n+1}{n-1} \cdot t_{\textrm{ch}}/\tau_0^{\textrm{MD}})$ and approximate the luminosity as a constant over time. In the opposite limit $t_{\textrm{ch}}/\tau_0 \gg 1$, the luminosity drops very fast and we can see the injection as a burst.


If the loss mechanism responsible for the spin-down were exclusively magnetic dipole (MD) emission, then the braking index would be $n = 3$ \cite{2005handbook} and the characteristic timescale of the frequency (and luminosity) drop would be given by $\displaystyle{\tau_0^{\textrm{MD}} = \frac{3 I c^3}{B^2 R^6 \Omega_0^2}}$, where $I$ is the moment of inertia of the spinning neutron star, $B$ is the surface magnetic field, $\Omega_0$ is the initial frequency. 

For all the nearby pulsars tabulated in the ATNF catalogue\footnote{\url{http://www.atnf.csiro.au/people/pulsar/psrcat/}} \cite{2005AJ....129.1993M}, the ratio $t_{\textrm{ch}}/\tau^{\textrm{MD}}_0$ given above is typically one order of magnitude lower than $1$ ($\sim 0.3$), which would point towards a constant-luminosity injection.

However, $n$ can be inferred only when observations are long enough to allow the derivation of all three quantities $\Omega, \dot{\Omega}, \ddot{\Omega}$. For this reason, they are available for a limited number of cases only \cite{2015PhRvD..91f3007H}, and in each of them the results show values of $1.9 < n < 2.8$, significantly different from the ideal MD model. Moreover, a comparison between the energy budget released by the pulsars calculated via MD-emission with the same quantity derived by observations, independently of the emission model, reveals significant discrepancies, as discussed in detail in Appendix \ref{app:appendix_pulsars_ATNF}. Finally, even if the constant-luminosity injection were a good approximation, it would become progressively more unreliable for increasing pulsar age.

For these reasons, we are led to conclude that other energy-loss mechanisms, rather than MD-emission only, might be at work. Thus, in the following we will consider only the model-independent equations \eqref{eq:energy_loss_pulsar}-\eqref{eq:decaying_luminosity} and study the two limiting cases of burst-like (discussed many times in the literature) and constant-luminosity injection of $e^{\pm}$, in order to bracket the above-mentioned uncertainty.

\item As far as the acceleration spectrum is concerned, we recall that the broad-band radiation emitted by PWNe can be typically modeled as synchrotron and IC emission from a population of relativistic electrons and positrons distributed in energy as a broken power-law. These leptonic pairs, initially extracted by the surface of the neutron star, are then most likely accelerated at, or close to, the termination shock (TS) by a variety of possible mechanisms.

The current data probing the non-thermal radiation (in \textit{Radio} and \textit{X-ray} frequencies) emitted from several well-observed PWNe \cite{Jankowski:2017yje} require a lepton spectrum which has the shape of a broken power law, with a hard spectrum (with slope $1 \lesssim \Gamma_{\textrm{inj}} \lesssim 2 $) below a break at $\sim 200$ -- $400$ GeV, and a steeper one ($\Gamma_{\textrm{inj}} > 2$) at larger energies (see \cite{Bykov:2017xpo,2011ASSP...21..624B,Amato:2013fua,Bucciantini:2010pd}). The hard, low-energy spectrum has been object of debate over the years, and several acceleration mechanisms were proposed, including  magnetic
reconnection at the TS and resonant absorption of ion-cyclotron waves.

Motivated by these considerations, in the following we will adopt both a broken power-law and a single power law with exponential cutoff and compare our result with those obtained in several previous analyses (see for instance the recent reviews \cite{Gaggero:2018zbd,Gabici:2019jvz} and the references therein).

\end{enumerate}

As a final remark, we point out that the particles are expected to be released from the PWN region with some delay. A minimal contribution to this delay is given by the time the pulsar --- due to its proper motions --- takes to leave the associate SNR shell, which we estimate to be $t_{\textrm{rel}} = 6.4 \cdot 10^{4}$ yr for pulsars (see Appendix \ref{app:release_time_from_PWN}).
That estimate could be even larger if we were to take into account the results of recent analyses of the HAWC \cite{2017Sci...358..911A} and Fermi-LAT \cite{2019arXiv190305647D} data for the Geminga and Monogem regions, showing that $e^\pm$ diffusion may be even more delayed around those objects. However, the possible consequences of these pockets of slow diffusion (commonly called \textit{TeV halos}) around PWNe still have to be determined. In fact, while \cite{2018PhRvD..97l3008P}, for instance, states that a two-zone model separating the TeV halo from the rest of the ISM still allows positrons from Geminga and Monogem to reach the Earth, in another recent study \cite{Johannesson:2019jlk} the authors argue that the same result depends on the size and other properties of the halo. We believe that the growing interest of the community in these TeV halos will lead to dedicated observations of other similar high-confinement regions, in order to establish if they are present around each PWN, as already outlined in \cite{Linden:2017vvb}. Collecting more statistics will eventually allow to infer their physical properties and to shed light on the puzzle of the positron origin.

\subsection{Diffusive propagation of leptons in the Galaxy: study of the analytical solution}\label{subsec:study_analytical_solution_pulsars}

With the parametrization of the source term and the delay of the particle release properly settled, we now turn our attention to the propagation of the electron+positron pairs from individual sources in the ISM.

We describe the transport process by means of a simplified version of Equation \eqref{eq:prop}, where low-energy effects such as advection and reacceleration are neglected and spherical symmetry is assumed:
\begin{equation}\label{eq:transport_equation_polar_coordinates}
    \frac{\partial f(E,t,r)}{\partial t} = \frac{D(E)}{r^2} \frac{\partial}{\partial r} r^2 \frac{\partial f}{\partial r} + \frac{\partial}{\partial E} (b(E) f) + Q,
\end{equation}
where $b(E)$ is the energy-loss rate. This term, in general, takes into account a variety of processes: ionization, Coulomb scattering, bremsstrahlung, Inverse Compton, synchrotron. 
Whereas the {\tt DRAGON} setup properly accounts for all of them, in this section we approximate $b(E)$ as
\begin{equation}\label{eq:energy_losses_leptons}
\frac{dE}{dt} \simeq - b_0 \, E^{2}
\end{equation}
with $b_0 = 1.4 \cdot 10^{-16} \, \text{GeV}^{-1} \, \text{s}^{-1}$, corresponding to a typical local interstellar gas density of $1 \, \text{cm}^{-3}$ and a total magnetic field  $B_{\textrm{tot}} = 5 \, \mu G$, compatible with a recent analysis~\cite{Sofue_2019}. This expression captures the dominant leptonic processes (Inverse Compton and synchrotron) in the local environment, as far as the energy range of interest for the present work is concerned ($E > 1$ GeV). It is worth mentioning that, although a full numerical treatment of the energy losses for relativistic leptons would require a correction to the $\propto E^2$ scaling due to the Klein-Nishina calculation of the IC scattering~\cite{Blumenthal:1970gc}, the authors of \cite{refId0} showed that the propagated spectra would change only up to a factor of $\sim 1.5$ in normalization for the adopted value of $B_{\textrm{tot}}$ (see their Figure 2). This uncertainty does not significantly affect the conclusions of this work, therefore we neglect the full treatment.

\begin{figure}
    \begin{subfigure}{.5\linewidth}
    \centering
        \includegraphics[width=1.\linewidth]{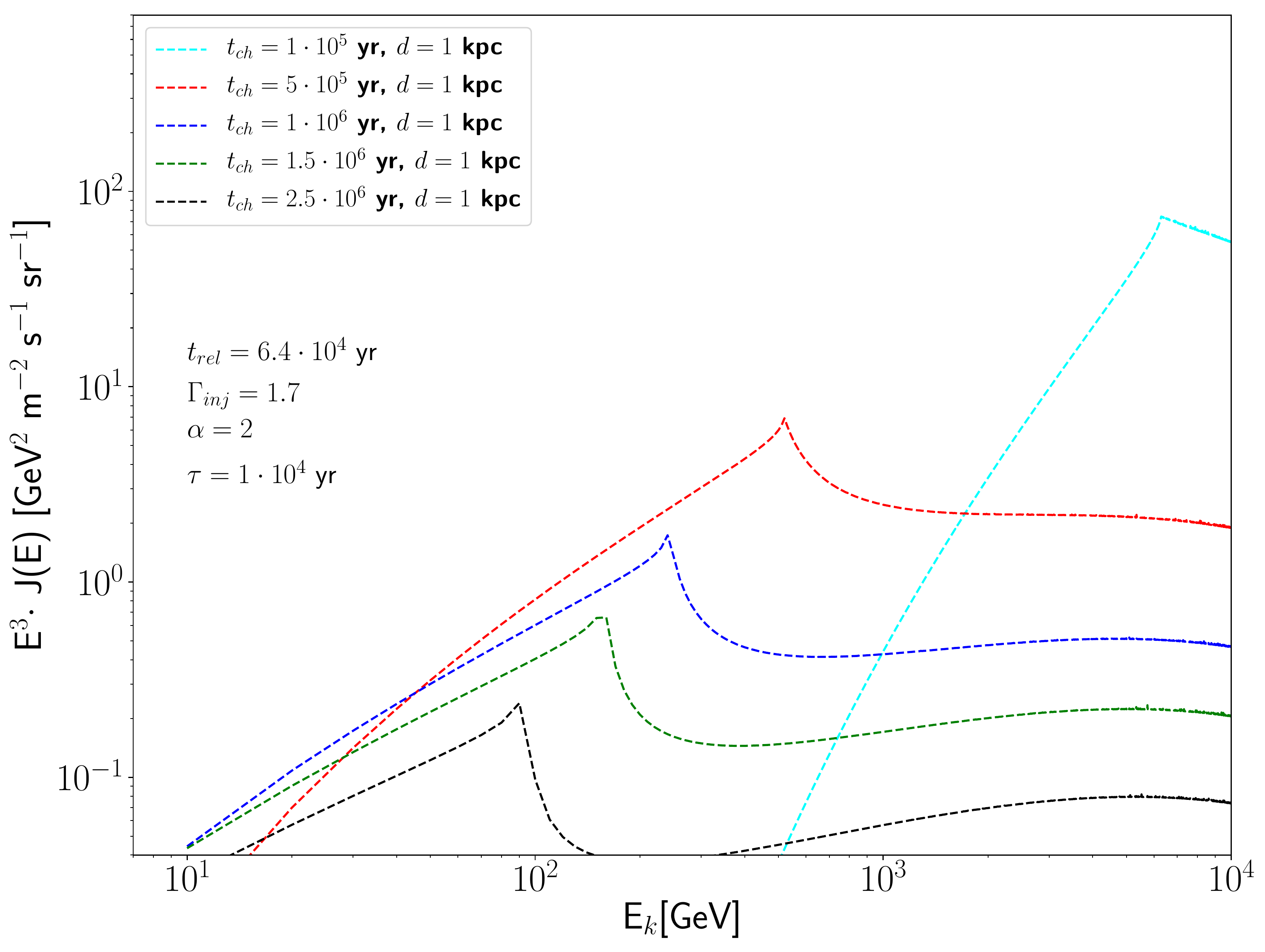}
        \caption{}
        \label{subfig:pulsar_extra_component_decaying_L_hard_inj_1kpc}
    \end{subfigure}
    \begin{subfigure}{.5\linewidth}
    \centering
        \includegraphics[width=1.0\linewidth]{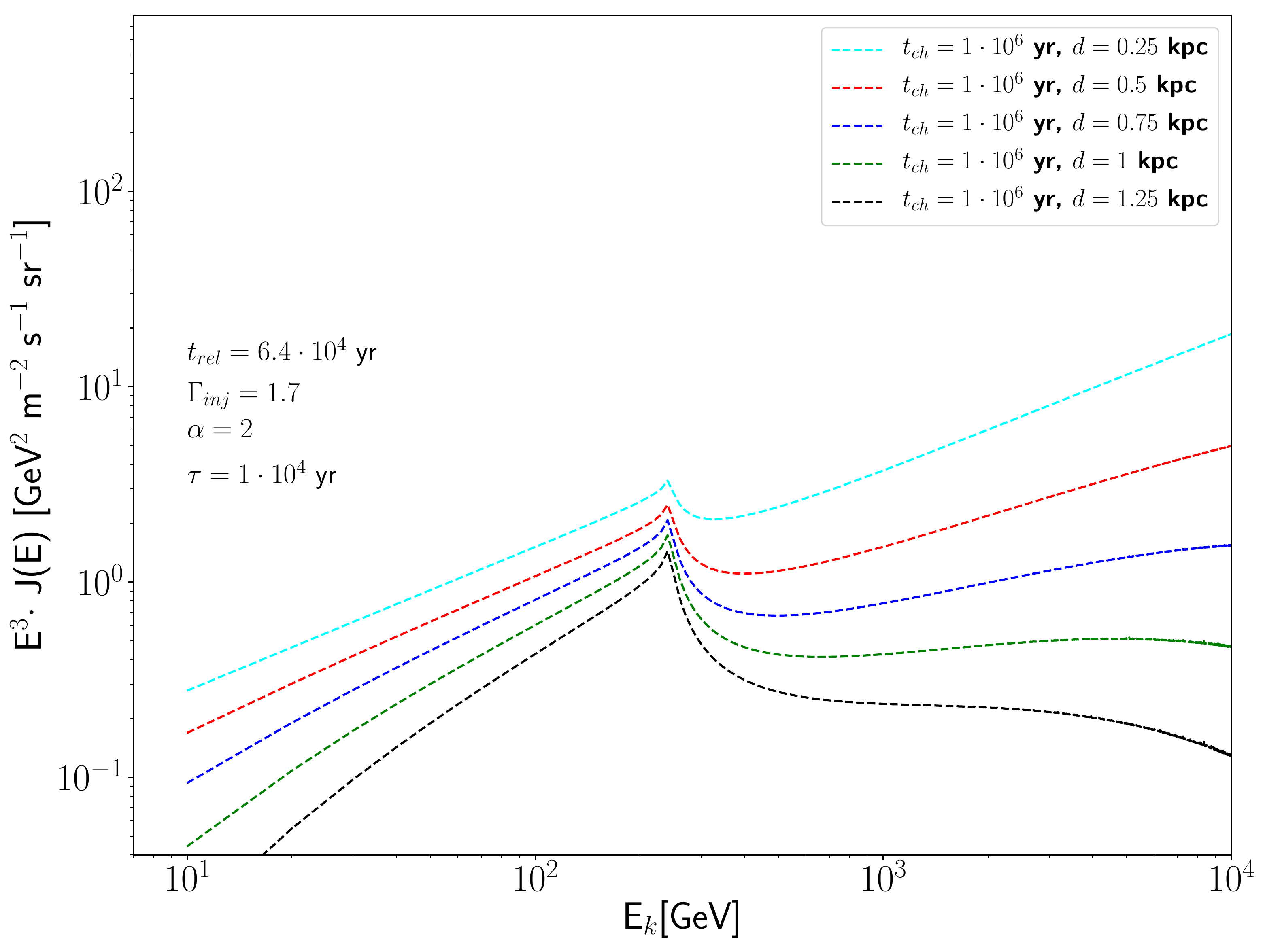}
        \caption{}
        \label{subfig:pulsar_extra_component_decaying_L_hard_inj_different_distance}
    \end{subfigure}
    \caption{\small{Solution of the transport equation for a decaying-luminosity single source, plotted for objects of (a) different ages and fixed distance (1 kpc) and (b) different distances and fixed age ($1 \cdot 10^6$ yr). The order of magnitude of the energy-budget is compatible with the one expected from pulsar emission ($\mathcal{O}(10^{47}-10^{49} \textrm{erg})$). The injection index is $\Gamma_{\textrm{inj}} = 1.7$, although we verified that the shifting is independent of it. As the source age increases, the emission peak shifts to the low-energy range.}}
    \label{fig:pulsar_solution_study_decaying_L}
\end{figure}

Equation \eqref{eq:transport_equation_polar_coordinates} can be solved analytically \cite{1995PhRvD..52.3265A}, as detailed in Appendix \ref{app:solution_transport_equation}. 
For the purpose of this work, we are interested in the behaviour of the solution as a function of the age and the distance. A time-decaying luminosity function as given in Equation \eqref{eq:decaying_luminosity}, assuming a power-law injection spectrum, yields the solutions plotted in  
Figure \ref{fig:pulsar_solution_study_decaying_L}. 

The prominent peak in the solution is due (at fixed distance) to the interplay between the diffusion dominating at low energy and the energy losses at high energy. While a burst-like injection gives rise to a sharp cutoff above the peak energy, a long-lasting source results in a plateau or even a growing-with-energy behaviour for large values of $\tau_0$ or short distances. 

Taking into account the possible presence of a UV cutoff in the source spectrum (see discussion in the previous paragraph), the peak energy is determined by the condition
\begin{equation}\label{eq:maximum_energy_losses}
    E_{\textrm{max}}(t) = \textrm{min} \left\{ 
    \frac{1}{b_0 (t - t_{\textrm{rel}})},  E_{\textrm{cut}} \right\},
\end{equation}
where $t$ is the age of the source and $t_{\textrm{rel}}$ the time it takes for particles to leave the source region. 
Therefore, the peak progressively shifts towards lower energies for increasing PWN ages.

\subsection{The contribution from old and young pulsars to the positron flux}\label{subsec:lepton_component_PWNe}

We start by considering the low-energy part of the positron spectrum and assume that it is originated by a large number of PWNe with age older than $\sim 10^6$ years.

This assumption is motivated by the trend of the peak energy outlined above and by the fact that,
below $\sim 100$ GeV, the diffusion horizon ($d_{\rm max} = \sqrt{4D(E) (t - t_{\textrm{rel}})}$) 
grows up to few kiloparsecs. Within that distance, a very large number of pulsars are observed, and --- provided that the diffusive time of their injected particles is smaller than their ages --- all of them are expected to contribute to the flux reaching the Earth, at energies that get lower with increasing age, as already outlined in \cite{refId0}. The cumulative spectrum of this ``large scale" $e^\pm$ component is therefore the convolution of the contributions from many single sources, integrated over their age distribution. 

A detailed Monte Carlo simulation of this integrated spectrum is beyond the scope of this paper and is postponed to a dedicated work. However, we tested the cumulative contribution from a sample of $10^4$ pulsars with ages between $10^6$ and $10^8$ yr (the sample number is compatible with the observed SN rate \cite{1999A&A...351..459C}), assuming that $e^\pm$ pairs are injected from these sources with a total energy budget in the [$10^{46}$ -- $10^{49}$] erg range, and with spectral indices between $1.3$ and $1.9$. We found that the simulated total spectrum from those sources displays a small scatter for different realizations of the pulsar distribution and --- with good approximation --- typically follows a smooth power-law.

Motivated by these considerations, we choose to consider an effective modeling of such large scale $e^\pm$ component within the {\tt DRAGON} framework, similarly to what done in previous works (see \textit{e.g.} \cite{2013JCAP...03..036D}).
Therefore, we add to our setup a charge-symmetric smooth \textit{extra-component} with the same spatial distribution of SNRs and tune its normalization and slope ($\Gamma_{\textrm{extra}} = 2.28$) to reproduce the AMS-02 data. It is important to remark that, given the large number of sources involved, the resulting convoluted soft ($\Gamma_{\textrm{extra}} > 2$) spectrum is not related to each single-source hard ($\Gamma_{\textrm{inj}} < 2$) injection.

We now focus on the high-energy part $E > 100$ GeV of the $e^\pm$ spectrum which should receive a significant contribution either from relatively young pulsars ($t \lesssim 10^5$ years) or even by older pulsars if they are long lived.

The key aspect in this energy domain is the pronounced drop-off in the positron spectrum observed by AMS-02 above $\sim 250$ GeV. 
The considerations discussed so far may lead us to two distinct interpretations of this feature:

\begin{itemize}
    \item Given the properties of the analytical solution, assuming that no relevant spectral steepening or cutoff is present at the source in this energy range, it is possible to ascribe the feature to the interplay between diffusion and energy loss. This would imply a dominant contribution in this range from a number of pulsar wind nebulae of approximate age of $\sim 10^6$ yr (see Figure \ref{subfig:pulsar_extra_component_decaying_L_hard_inj_1kpc}). Besides, in order to reproduce the above-mentioned drop-off in the data, such PWNe should be at a distance larger than or similar to $\sim \sqrt{4 \, D(E=230 \textrm{ GeV}) \cdot (t_{\textrm{age}} = 10^6 \textrm{ yr})} \approx 1.5$ kpc (see Figure \ref{subfig:pulsar_extra_component_decaying_L_hard_inj_different_distance}).
    \item Alternatively, given our knowledge of the injection spectrum of PWNe, summarized in Section \ref{subsec:Setting_the_stage}, a natural interpretation is that the positron flux around $200$ GeV is dominated by few (or one) nearby, young pulsar wind nebulae, which provide a relevant contribution on top of the diffuse, large-scale component discussed above, and is characterized by either a spectral break or a cutoff at that energy, explained by the acceleration processes taking place near the termination shock.
\end{itemize}

In what follows we will explore the second option leaving a deeper analysis of the first one to a forthcoming work. We just mention that a detailed Monte Carlo simulations was recently performed in \cite{2018PhRvD..98f3008C}\footnote{Interestingly, their model E1 --- which is characterized by diffusion and loss parameters very close to those adopted in this work --- predicts a positron fraction steadily growing with energy up to 100 GeV; above that energy, the fraction flattens reaching a maximum at about 300 GeV.}.

\subsection{Characterization of the high-energy flux}\label{subsec:bayesian_fit_positrons}

We here investigate in further details the case where, on top of the secondary positron flux and a large-scale extra component associated to a large number of old PWNe --- as discussed in Section \ref{subsec:lepton_component_PWNe} --- the high-energy positron flux is dominated by the contribution from a prominent young object featuring a break or a cutoff in the injection spectrum of $e^{\pm}$ pairs. 

In order to do so, we consider four different scenarios, deriving from the combination of two limit behaviours of the luminosity function (\textit{i.e.} burst-like injection and constant-luminosity injection) with the two possibilities for the injection feature (\textit{i.e.} exponential cutoff and break).

These are parametrized in the single-source term $Q(E,r,t)$ of the transport equation \eqref{eq:transport_equation_polar_coordinates} (for a detailed discussion on the solutions here used, we refer again to Appendix \ref{app:solution_transport_equation}).

In each case, the properties of the young, dominant object are assessed by means of a Bayesian fit. We consider data from AMS-02 \cite{AMS2019PhRvL.122d1102A} from 20 GeV on, to avoid problems deriving from solar modulation. We implement our theoretical knowledge of the problem by setting priors on the injection index, that we expect to be $\Gamma_{\textrm{inj}} \in [1,2]$, and on the critical energy above which we expect the injection feature to come into play, $E_{\textrm{cut, break}} > 150$ GeV. For the burst-like injection we consider the age and distance of the Monogem pulsar, while for the constant-luminosity we use the age and distance of Geminga. This is in accordance to what is shown and discussed in Appendix \ref{app:appendix_pulsars_ATNF}, where all the high-energy nearby (within $1.3$ kpc) sources are plotted in both injection scenarios, and the dominant contribution is assessed in both cases.

\begin{figure}
    \begin{subfigure}{.5\linewidth}
        \centering
        \includegraphics[width=1.02\linewidth]{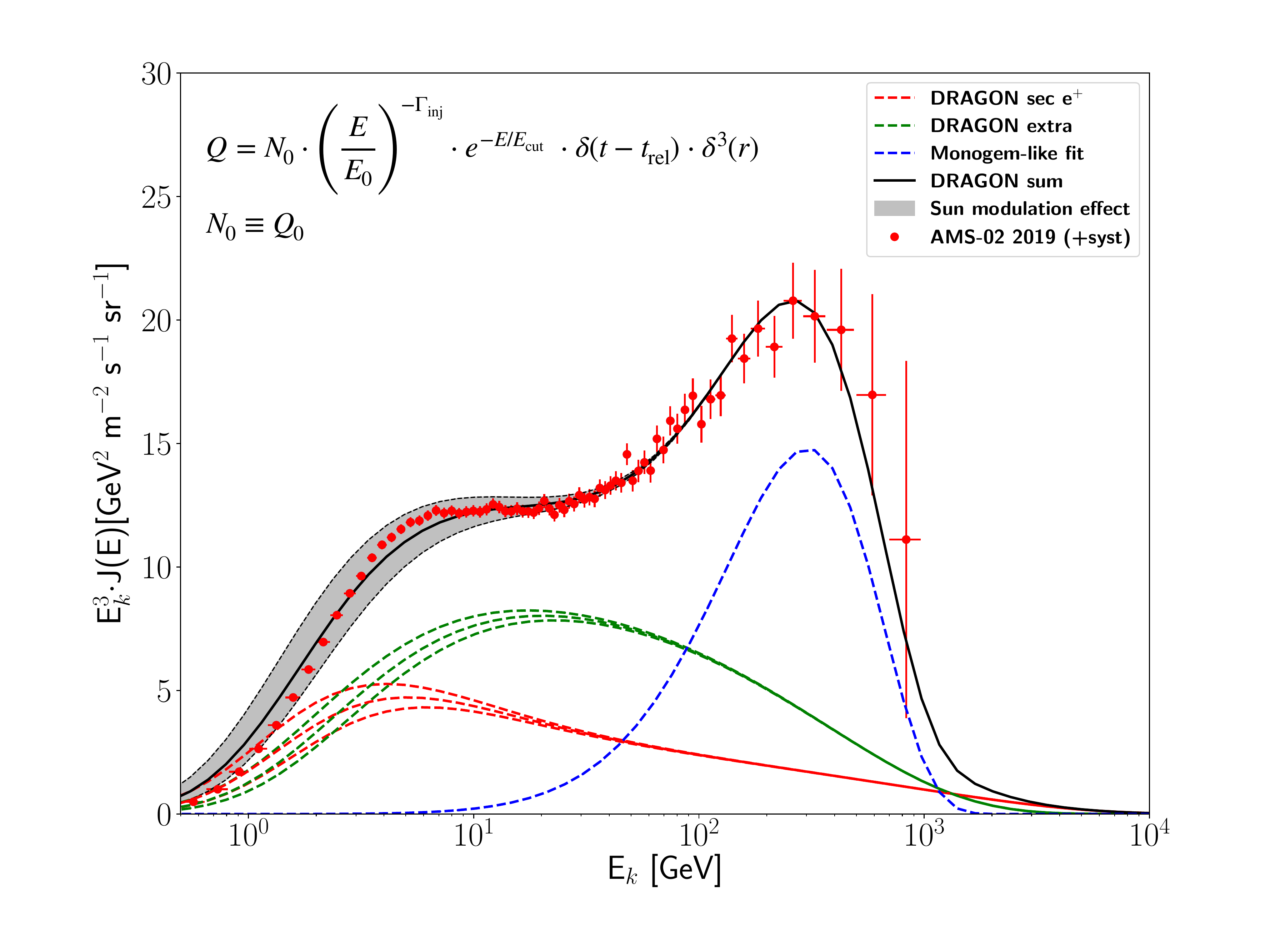}
        \caption{}
        \label{fig:positron_flux_1_source_burst_cutoff}
    \end{subfigure}
    \begin{subfigure}{.5\linewidth}
        \centering
        \includegraphics[width=1.02\linewidth]{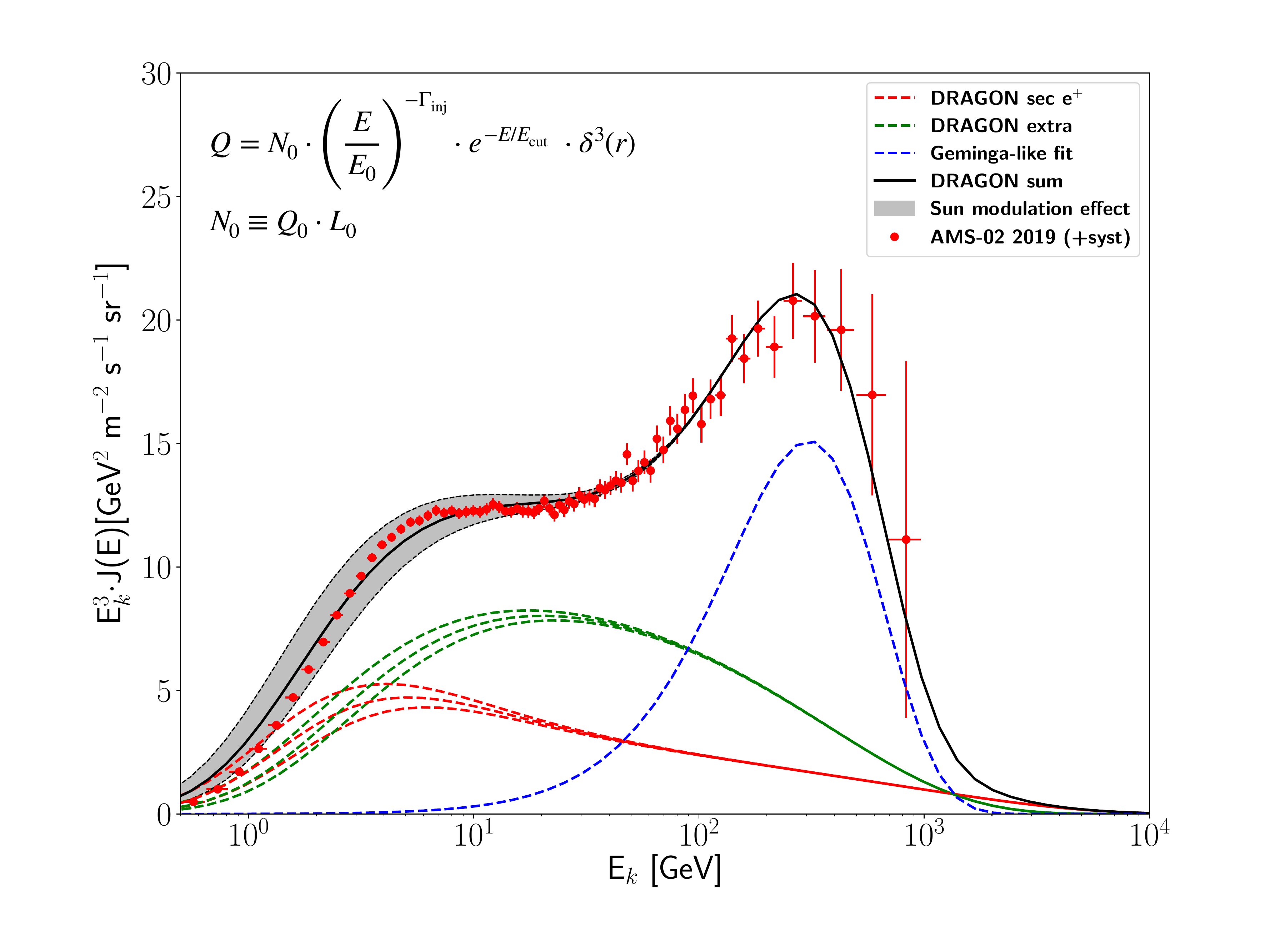}
        \caption{}
        \label{fig:positron_flux_1_source_continuous_L_cutoff}
    \end{subfigure}
    \begin{subfigure}{.5\linewidth}
        \centering
        \includegraphics[width=1.02\linewidth]{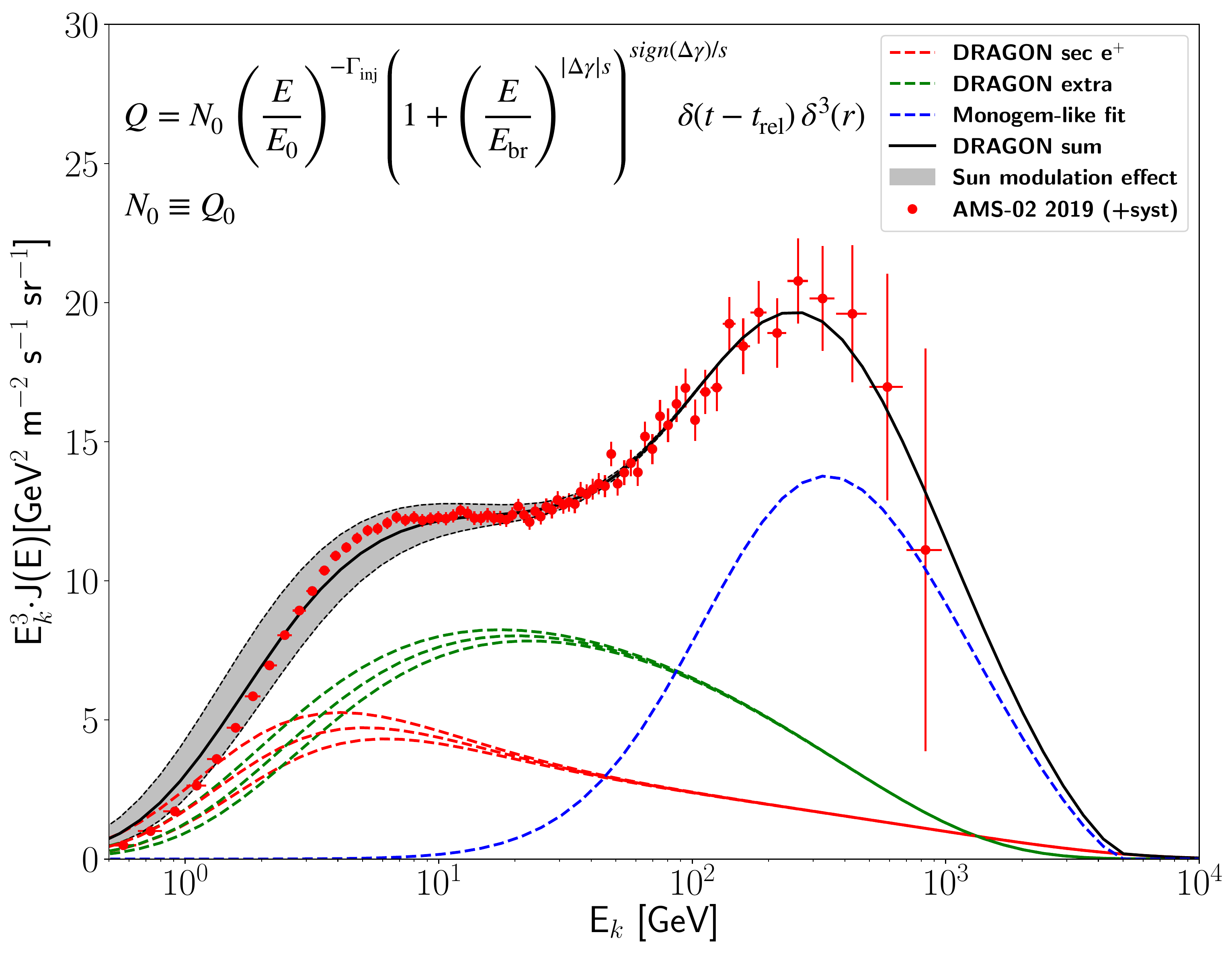}
        \caption{}
        \label{fig:positron_flux_1_source_burst_break}
    \end{subfigure}
    \begin{subfigure}{.5\linewidth}
        \centering
        \includegraphics[width=1.02\linewidth]{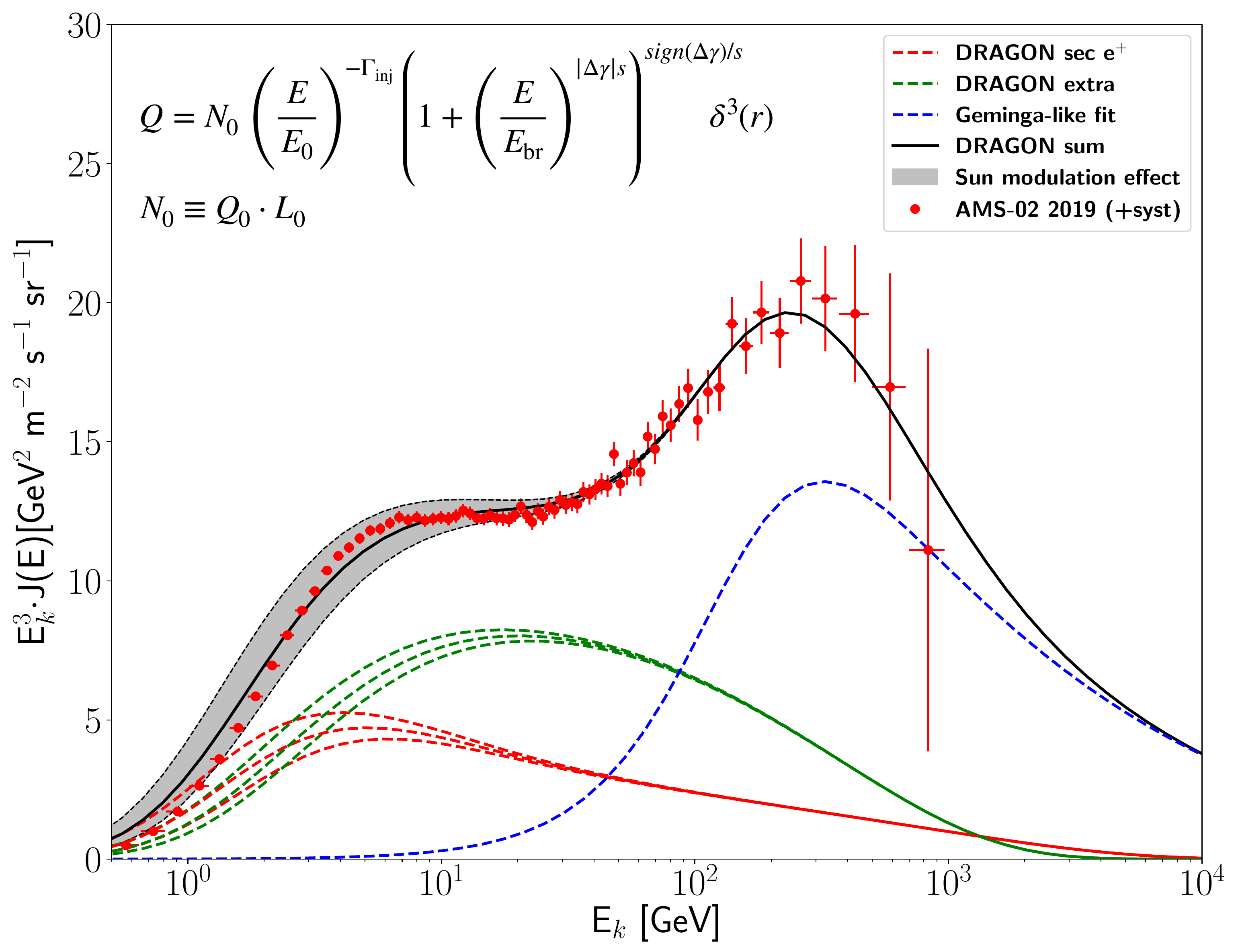}
        \caption{}
        \label{fig:positron_flux_AMS_1_source_constant_L_break}
    \end{subfigure}
\caption{\small{Fit to the positron flux for two classes of injection scenarios, where intrinsic features are added. (a) Burst-like injection with cutoff, (b) constant-luminosity injection with cutoff, (c) burst-like injection with broken power-law, (d) constant-luminosity injection with a broken power-law.}}
\label{fig:bayesian_fit_positron_flux_1_source}
\end{figure}

The resulting plots are shown in Figure \ref{fig:bayesian_fit_positron_flux_1_source}, where the source terms entering each fit function are shown inside each canvas, and the \textit{Maximum-a-Posteriori} (MAP) parameters of the fits --- (\textit{i.e.} the maximum values of the posterior distribution functions obtained as an output in the fitting procedure)--- are listed in Tables \ref{tab:positron_flux_best_fit_cutoff} and
\ref{tab:positron_flux_best_fit_break}. 

\begin{table}[t]
    \centering
    \begin{tabular}{|c|c|c|c||c|c|}
    \hline
        & $\bm{N_0}$ & $\bm{\Gamma_{\textrm{inj}}}$ & $\bm{E_{\textrm{cut}}}$ [GeV] & $\bm{E_{\textrm{tot}}}$ [erg] & $\bm{\eta^{\pm}}$ \\
         \hline 
         \hline
        \textbf{Burst} & $2.4 \cdot 10^{48} \, [\textrm{GeV}]^{-1}$ & $1.31$ & $270.78$ & $ 2 \cdot 5.39 \cdot 10^{46}$ & 0.8 \\
         \hline
        $\bm{L_0}$ & $1.17 \cdot 10^{35} \, [\textrm{GeV} \cdot \textrm{s}]^{-1}$  & $1.07$ & $200.43$ & $2 \cdot 2.02 \cdot 10^{45}$ & $< 1.2 \cdot 10^{-2}$ \\ 
        \hline
    \end{tabular}
    \caption{\small{Our MAP values for the injection parameters from $e^{\pm}$ sources with an intrinsic cutoff, set to have a prior distribution with $E_{\textrm{cut}} > 150$ GeV. The total energy injected in the ISM in the form of leptons is indirectly computed from the fit-parameters: the factor $2$ is multiplied because of the $e^{\pm}$ symmetry. The conversion efficiency $\eta^{\pm}$ is calculated with respect to the nominal ATNF observed parameters: for what explained in the text, this is an upper bound.}}
    \label{tab:positron_flux_best_fit_cutoff}
\end{table}

\begin{table}
    \centering
    \begin{tabular}{|c|c|c|c|c|c||c|c|}
    \hline
        & $\bm{N_0}$ & $\bm{\Gamma_{\textrm{inj}}}$ & $\bm{\Delta \gamma}$ & $\bm{E_{\textrm{break}}}$ [GeV] & $\bm{s}$ & $\bm{E_{\textrm{tot}}}$ [erg] & $\bm{\eta^{\pm}}$ \\
         \hline 
         \hline
        \textbf{Burst} & $1.08 \cdot 10^{48}\, [\textrm{GeV}]^{-1}$ & $1.02$ & $-2.77$ & $321.65$ & $0.31$ & $2 \cdot 2.35 \cdot 10^{47}$ & $\mathcal{O}(1)$ \\
         \hline
        $\bm{L_0}$ & $ 1.11 \cdot 10^{35} \, [\textrm{GeV} \cdot \textrm{s}]^{-1}$ & $1.10$ & $-1.74$ & $158.02$ & $1.11$ & $2 \cdot 3.35 \cdot 10^{47}$ & $\mathcal{O}(1)$ \\
        \hline
    \end{tabular}
    \caption{\small{Our MAP values for the injection-parameters from $e^{\pm}$ sources with an injection break, parametrized by the multiplying factor $\left( 1 + \left( \frac{E}{E_{\textrm{break}}} \right)^{|\Delta \gamma| \cdot s} \right)^{ \textrm{sign}(\Delta \gamma)/s}$, set to have a prior distribution with $E_{\textrm{break}} > 150$ GeV. The total energy injected in the ISM in the form of leptons is indirectly computed from the fit-parameters: the factor $2$ is multiplied because of the $e^{\pm}$ symmetry. The conversion efficiency $\eta^{\pm}$ is calculated with respect to the nominal ATNF observed parameters: for what explained in the text, this is an upper bound.}}
    \label{tab:positron_flux_best_fit_break}
\end{table}

We notice that each of the four combinations is compatible with the positron data. 
Nonetheless, comparing the numerical values on the tables, relevant physical aspects have to be noticed:
\begin{itemize}
    \item Even though we set a prior for the injection indices to be hard, data seem to favorite the very-hard end of the range: all the cases present $\Gamma_{\textrm{inj}} \lesssim 1.3$, with the softest being the burst-like injection with intrinsic cutoff. 
    \item For the burst-like solutions the injection features are found at energies higher ($E_{\textrm{cut,break}} > 270$ GeV) with respect to the constant-luminosity case ($E_{\textrm{cut,break}} \lesssim 200$ GeV): this effect is due to the peculiar shape of the burst-like solution, which features a sharp cutoff that is required to match the drop-off of the data. 
    \item The total amount of energy converted into $e^{\pm}$ pairs is estimated by means of: 
    \begin{equation}\label{eq:total_energy_injected}
        E_{\textrm{tot}} = \int_{E_{\textrm{min}}}^{E_{\textrm{max}}} dE \int_{t_{\textrm{rel}}}^{t_{\textrm{age}}} dt \int d^3 \vec{r} \, E \cdot Q(E,\vec{r},t),
    \end{equation}
    where $E_{\textrm{min}} = 1$ GeV and $E_{\textrm{max}} = +\infty$: only in the cases of logarithmic divergences a cut at very high-energies ($E_{\textrm{cut}} = 100$ TeV) is set. Equation \eqref{eq:total_energy_injected} gives values compatible with the order-of-magnitude energies that are thought to be injected by pulsars in the ISM. Besides, an efficiency is estimated with respect to the total energy injected by the source, that we compute multiplying the observed rate of rotational-energy loss $\dot{E}_{\textrm{rot}} = \frac{d}{dt} \left(  \frac{1}{2} I \Omega^2 \right) = - I \Omega \dot{\Omega}$ by the characteristic age of the source. As discussed in detail in Appendix \ref{app:appendix_pulsars_ATNF}, we observe that the quantity thus computed is actually a lower bound. Nevertheless, we notice that the values of $\eta^{\pm}$ estimated in the two injection scenarios are very different. This is not unexpected: in fact, at given age $t_{\textrm{ch}}$ and loss rate $\dot{E}_{\textrm{rot}}$, if a source is continuously emitting, then the total amount of energy injected in the ISM is much larger than in the burst-like case. Therefore, to match with the observed lepton spectrum, only a much smaller fraction of this energy needs to be converted into leptons \cite{2011ASSP...21..624B}. These observations are visible only in the cases with a cutoff in the injection and are compatible with what is shown in Appendix \ref{app:appendix_pulsars_ATNF}.
\end{itemize}

We point out that energetics (as listed in Table \ref{tab:positron_flux_best_fit_break}) cannot be taken as a strong argument against one scenario or the other, because we do not have a better model-independent estimation for $E_{\textrm{tot}}$, and also because of the large statistical uncertainties on the high-energy positron flux. Future data with more statistics and higher energies may play a crucial role in this context: for instance, an additional data point in the TeV domain may allow to disentangle between the scenarios presented in the upper and lower panels of Figure \ref{fig:bayesian_fit_positron_flux_1_source}.

As a final comment, we also remark that the grey band accounting for the solar modulation is within the intervals identified by the time structures discussed in \cite{Aguilar:2018ons}.

In conclusion, in this section we found that scenarios characterized by a prominent young pulsar that dominates the high-energy positron flux, and a large number of middle-aged and old pulsars --- modeled as a continuous contribution to the flux --- are compatible with current data, under different hypotheses on both the injection spectrum and the timescale of the luminosity decline. The best-fit values for the injection spectra are compatible with the physical mechanisms outlined at the beginning of the section. However, different scenarios correspond to different estimates of the total energy budget and to a different hierarchy of the contributions from the nearby pulsars, as shown in Figure \ref{fig:pulsars_from_ATNF_HE}. Therefore, given the current data and the current knowledge on the physics of pulsar wind nebula emission, it is not possible to clearly identify which objects actually provide the most relevant contribution to the positrons. Nevertheless, the measurement of the absolute positron flux has important implications. In fact, positrons are likely emitted in $e^{\pm}$ pairs, giving us the exact contribution of this class of sources to the electron flux as well. Therefore, since a significant part of the $e^{+} + e^{-}$ spectrum is still missing after accounting for these contributions, we state that the high-energy lepton flux requires the presence of a different class of local electron-only sources. This allows us to focus on the next section, without worrying about the uncertainties on the positron origin.

\section{Local electron accelerators explain the high-energy electron data}\label{sec:local_SNR}

This section is dedicated to the interpretation of the all-lepton spectrum. We adopt the best-fit CR transport scenario evaluated in Section \ref{sec:characterizazione_CR_transport} and the best-fit $e^\pm$ flux (assumed charge symmetric) determined in the previous section for one of the four combinations discussed: the specific choice for the pulsar injection setup does not affect the results presented in this section.
We will show that the closest observed SNRs are not sufficient to describe the observed spectrum and an additional source with specific characteristics has to be invoked to reproduce in particular the $\sim 1$ TeV break recently measured by the space-born and ground-based experiments H.E.S.S., VERITAS, CALET and DAMPE. Even though no information is given on the nature of the object, we model it as a SNR. This is because, mainly based on energetic arguments (see for instance \cite{Blasi:2013rva}), these objects are expected to provide the bulk of CRs observed at the Earth. Although not used here, we also notice that the combined study of the all-electron and radio emission of nearby SNR can also provide valuable complementary information   
(see \textit{e.g.} the recent \cite{2019JCAP...04..024M}). 

\subsection{Contribution from the known objects}

Multi-wavelength observations show the presence of five Supernova Remnants (SNRs) in the local region (within $\sim 1$ kpc) surrounding the Earth\footnote{\url{http://www.physics.umanitoba.ca/snr/SNRcat}} \cite{2012AdSpR..49.1313F}, identified with the names Vela Jr, Vela, Cygnus Loop, Simeis-147, IC-443.

We report in Table \ref{tab:SNR_list_parameters} the nominal ages and distances of these objects and the distances that particles with energy 1 TeV and 10 TeV can travel in the ISM via diffusive transport, as well as the ratios between the diffusive distance and the true distance of each source. 
We outline that, given the values reported in that table, the contribution of Vela Jr --- the youngest remnant in the set under consideration --- should peak around $\sim 100$ TeV, where we do not have reliable data. As far as the other SNRs are concerned, Vela is expected to provide the dominant contribution; the emissions of the other SNRs are expected to be subdominant, though not negligible, since their diffusive distance is smaller than or comparable to the nominal one. Therefore, we choose to take into account all the remnants listed above with the only exception of Vela Jr.

\begin{table}[t]
    \centering
    \begin{tabular}{|c|c|c|c|c|c|c|}
    \hline
        & $\bm{t}_{\textrm{age}}$ [yr] & $\bm{d}$ [pc] & $\bm{r_{\textrm{diff,1 TeV}}}$ [pc] & $\bm{r_{\textrm{diff,10 TeV}}}$ [pc] & $\frac{\bm{r_{\textrm{diff,1 TeV}}}}{\bm{d}}$ & $\frac{\bm{r_{\textrm{diff,10 TeV}}}}{\bm{d}}$ \\
         \hline 
         \hline
        \textbf{Vela Jr} & $2.5 \cdot 10^3$  & $214.2$ & $1.08 \cdot 10^2$ & $1.82 \cdot 10^2$ & $0.51$ & $0.85$ \\
         \hline
        \textbf{Vela} & $1.23 \cdot 10^4$  & $250.92$ & $2.69 \cdot 10^2$ & $4.52 \cdot 10^2$ & $1.07$ & $1.80$ \\
        \hline
        \textbf{Cygnus L} & $8 \cdot 10^3$  & $449.82$ & $2.17 \cdot 10^2$ & $3.64 \cdot 10^2$ & $0.48$ & $0.80$ \\
        \hline
        \textbf{Simeis-147} & $4 \cdot 10^4$  & $918$ & $4.85 \cdot 10^2$ & $8.14 \cdot 10^2$ & $0.52$ & $0.89$ \\
        \hline
        \textbf{IC-443} & $3 \cdot 10^4$ & $918$ & $4.20 \cdot 10^2$ & $7.05 \cdot 10^2$ & $0.46$ & $0.77$ \\
        \hline
    \end{tabular}
    \caption{\small{The nominal ages and distances of the five closest observed SNRs are listed. The diffusive distances are also shown for particles of 1 TeV and 10 TeV, in order to have a clear look on the sources that can contribute to the multi-TeV lepton flux. For a comparison with the loss-properties, $r_{\textrm{loss,1 TeV}} \simeq 1.15 \cdot 10^3$ pc and $r_{\textrm{loss,10 TeV}} \simeq 6.13 \cdot 10^2$ pc. From the numbers, Vela seems the one that can contribute the most to the $e^{+} + e^{-}$ flux.}}
    \label{tab:SNR_list_parameters}
\end{table}

In order to estimate the contributions from the sources mentioned above, we perform a fit, based on the all-lepton data from AMS-02 \cite{AMS2014PhRvL.113v1102A} plus H.E.S.S. \cite{kerszberg_ICRC}, in which each SNR is modeled as a continuous source of $e^{-}$. The choice of the AMS-02 data is consistent with the previous part of the analysis, since we calibrated our model based on AMS-02 observations. Then, we choose to consider H.E.S.S. data, although in their preliminary release, because they provide the best combination of up-to-date and highest-energy observations, and are consistent within the error band with AMS-02 experiment.

It is possible to parametrize the problem with the same formalism we used for the pulsar decaying-luminosity injection, \textit{i.e.} the luminosity function can be written as:
\begin{equation}\label{eq:luminosity_vs_time_SNR}
    L(t) = \frac{ L_0 }{ \left(  1 + \frac{t}{\tau_{\text{d}}}  \right)^{\alpha_{\text{d}}}},
\end{equation}
where now $\tau_{\textrm{d}}$ and $\alpha_{\textrm{d}}$ are specific for the release from a SNR and have nothing to do with pulsar injection mechanisms, and $t$ is as usual the age of the source. The particle propagation is accounted for by solving the transport equation as described in Appendix \ref{app:solution_transport_equation}.

The parameters we vary in the fitting procedure are the flux normalization, the injection index, and the luminosity-decline parameters ($\tau_{\textrm{d}}, \alpha_{\textrm{d}}$) of the sources. Based on the physical assumption that the acceleration mechanism is the \textit{diffusive shock acceleration} (DSA) \cite{1977SPhD...22..327K, bell1978, 1978ApJ...221L..29B, axford1977}, a prior is set for the injection indices to be $\Gamma_{\textrm{inj}} \in [2,3]$ \cite{2001RPPh...64..429M,Caprioli_2008}. The parameters ($\tau_{\textrm{d}}, \alpha_{\textrm{d}}$) are allowed to vary, but are set as identical for each source: we verified that this approximation has no significant impact on the final result, for values in the ranges $10^3 < \tau_{\textrm{d}} < 10^6 \textrm{ yr}$ and $1 < \alpha_{\textrm{d}} < 3$, due to the relatively large distance of the sources of interest.

\begin{figure}
    \centering
    \includegraphics[scale=0.4]{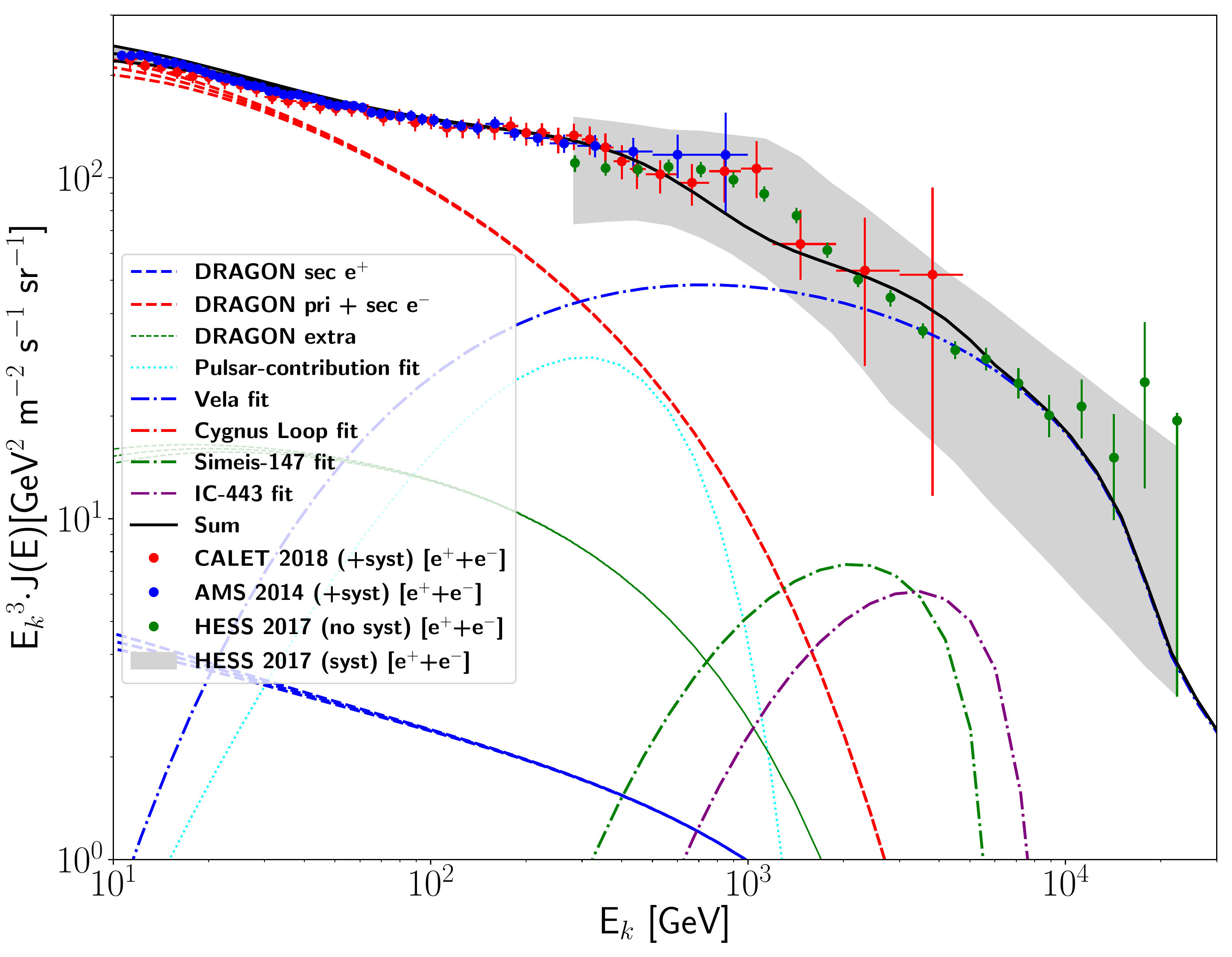}
    \caption{\small{Fit of the $e^{+}+e^{-}$ flux. The secondary and primary production and the extra-component, along with the fitted pulsar contribution, are considered as background, while the four SNRs have their parameters resulting from the fit. The blue dots are data from AMS-02 \cite{AMS2014PhRvL.113v1102A}, the red dots from CALET \cite{Adriani:2018ktz} and the green dots from H.E.S.S. \cite{kerszberg_ICRC}.}}
    \label{fig:all_electron_flux_4_SNR}
\end{figure}

The results are shown in Figure \ref{fig:all_electron_flux_4_SNR}, and the MAP parameters listed in Table \ref{tab:SNR_fit_results_parameters}.

\begin{table}[t]
    \centering
    \begin{tabular}{|c|c|c|c|c||c|}
    \hline
        & $\bm{N_0}$ $[\textrm{GeV} \cdot \textrm{s}]^{-1}$ & $\bm{\Gamma}_{\textrm{inj}}$ & $\bm{\tau}_{\textrm{d}}$ [yr] & $\bm{\alpha}_{\textrm{d}}$ & $\bm{E}_{\textrm{tot}}$ [erg] \\
        \hline 
        \hline
        \textbf{Vela} & $1.31 \cdot 10^{41}$ & $2.84$ & \multirow{4}{*}{$1.87 \cdot 10^3$} & \multirow{4}{*}{$2.47$} & $9.52 \cdot 10^{48}$ \\
        \cline{1-3} \cline{6-6}
        \textbf{Cygnus L} & $6.11 \cdot 10^{39}$ & $2.95$ &  &  & $3.78 \cdot 10^{47}$ \\
        \cline{1-3} \cline{6-6}
        \textbf{Simeis-147} & $3.98 \cdot 10^{42}$ & $2.98$ &  &  & $2.59 \cdot 10^{50}$ \\
        \cline{1-3} \cline{6-6}
        \textbf{IC-443} & $1.03 \cdot 10^{41}$ & $2.93$ &  &  & $7.04 \cdot 10^{48}$ \\
        \hline
    \end{tabular}
    \caption{\small{The table reports the MAP parameters resulting from the fit. Also, the total energy injected by each source in the form of $e^{\pm}$ is computed, based on the normalization.}}
    \label{tab:SNR_fit_results_parameters}
\end{table}

As expected, the main contribution comes from the Vela SNR, due to the interplay among the diffusive distance, the distance of the source and the energy-loss characteristic distance. Simeis-147 and IC-443 cannot give contribution to the $\mathcal{O}(10 \textrm{ TeV})$ flux, since their distance is larger than the loss distance at this energy, and indeed their peaks lie at energies smaller than $\sim 8$ TeV. The contribution from Cygnus Loop is extremely suppressed and even not visible in the plot, because the source is younger than the others and its peak would appear at an energy too-high to be compatible with the data. Finally, the energy budgets of those sources are compatible with those expected at SNR events ($\sim 10^{51}$ erg), taking into account the conversion efficiency into leptons within the range $\eta^{\pm} \sim 10^{-4}, ..., 10^{-1}$ \cite{Zirakashvili2017}, due to physical phenomena such as the particle escape at the shock front \cite{1978A&A....70..607S, Gabici:2011wn}.

The most relevant implication of this result is that the $\sim 1$ TeV spectral break cannot be reproduced with known sources. In fact, as noticed in \cite{Recchia:2018jun}, the propagated spectrum from a nearby SNR would peak at that energy only for a source as old as $\sim 2 \cdot 10^5$, a much larger age compared to that of the observed sources considered here. 

Finally, we notice that those conclusions strictly hold as long as we consider only statistical errors for the H.E.S.S. data. However, the systematic uncertainty quoted by the H.E.S.S. experiment is actually much larger than the statistical one, therefore the claim relies on the assumption of a very high correlation among the systematic errors of the different energy bins. We hope that a future better estimation of the covariance matrix will help to better assess the compatibility between this scenario and the data.

\subsection{Characterization of a potential source reproducing the $\sim 1$ TeV break}

A fit considering the emission of all the known sources in the current catalogs has shown that either a radical change in the propagation paradigm or an unknown source are needed. 
In particular, an old ($\sim 10^5$ yr) SNR seems to be necessary to reproduce correctly the $\sim 1$ TeV break, as first pointed out in \cite{Recchia:2018jun}.

In order to better characterize this potential source in terms of its distance and energy budget, we perform a fit of the data in two different scenarios:
\begin{enumerate}
    \item[1)] None of the listed known sources contribute to the flux,
    \item[2)] All of them add their maximal contributions to the flux.
\end{enumerate}

The free parameters in both cases are the normalization, the injection index, the ($\tau_{\textrm{d}},\alpha_{\textrm{d}}$) luminosity parameters, the age and distance of the source. We set a flat prior for the injection index in the range $\Gamma_{\textrm{inj}} \in [2,3]$, since we assume DSA to be the acceleration mechanism at work. In the second case, we also assume a flat prior for the distance in the range $d < 1.2 \cdot 10^3$ pc because we do not expect $\sim 1$ TeV leptons to come from more distant sources, due to energy-losses. For the fit, we use the same data set as before.

\begin{figure}
\centering
    \begin{subfigure}{.8\linewidth}
        \centering
        \includegraphics[scale=0.4]{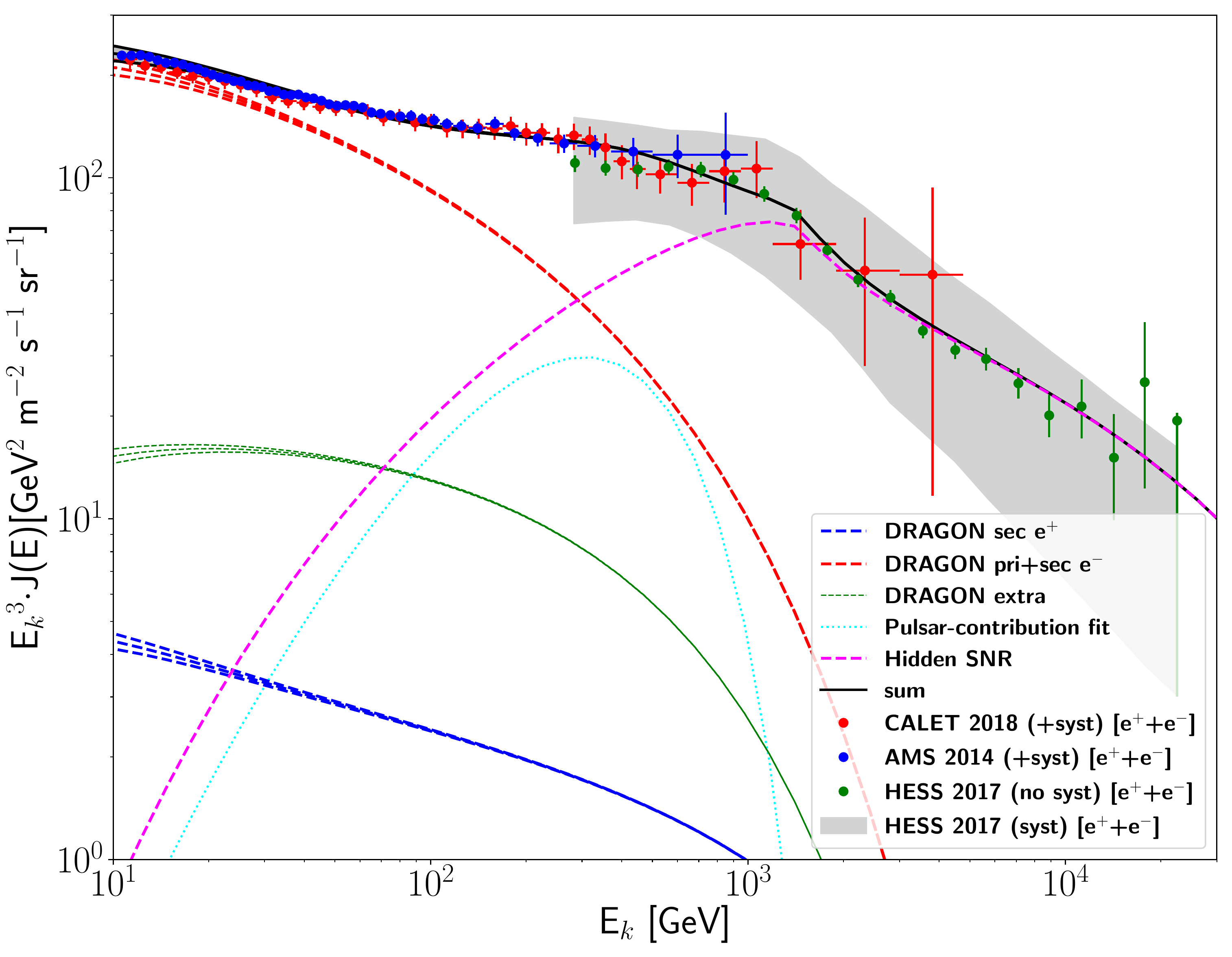}
        \caption{}
        \label{subfig:All_electron_flux_1_source}
    \end{subfigure}
    \begin{subfigure}{.8\linewidth}
        \centering
        \includegraphics[scale=0.4]{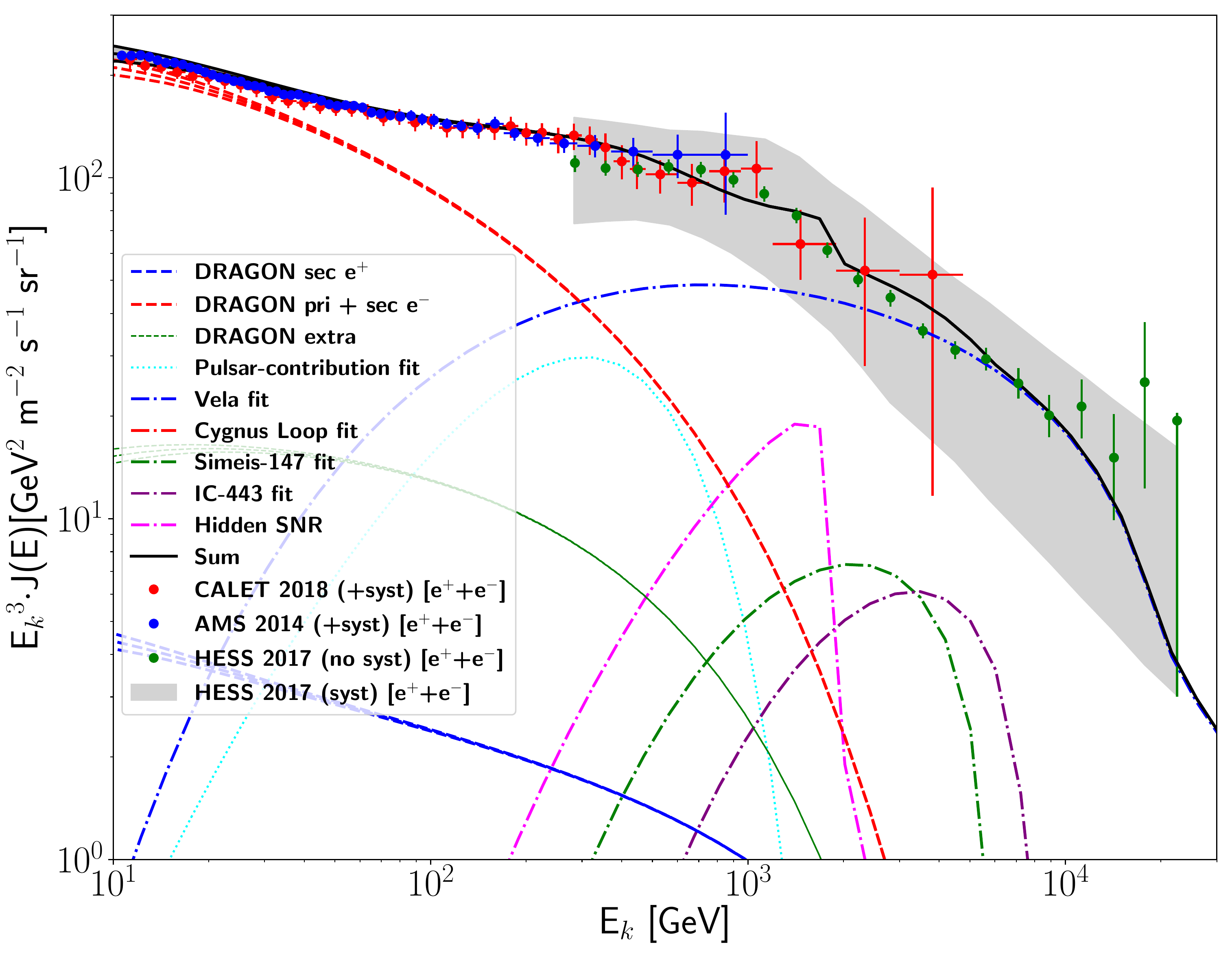}
        \caption{}
        \label{subfig:All_electron_flux_4_plus_1_SNR}
    \end{subfigure}
\caption{\small{Fits of the $e^{+}+e^{-}$ flux: the secondary and primary production, the extra-component and the fitted pulsar contribution, are considered as background. An additional hidden SNR with free parameters $N_0, \Gamma_{\textrm{inj}}, \alpha_{\textrm{d}}, \tau_{\textrm{d}}, t_{\textrm{age}}, r_{\textrm{dist}}$ is fitted when: (a) no known SNR is taken into account, (b) contributions from all the observed SNRs are also considered. The blue dots are data from AMS-02 \cite{AMS2014PhRvL.113v1102A}, the red dots from CALET \cite{Adriani:2018ktz} and the green dots from H.E.S.S. \cite{kerszberg_ICRC}.}}
\label{fig:All_electron_flux_fits_hidden_SNR}
\end{figure}

The outcome of this procedure is shown in 1) Figure \ref{subfig:All_electron_flux_1_source} and 2) Figure \ref{subfig:All_electron_flux_4_plus_1_SNR} and the parameters summarized in Table \ref{tab:SNR_fit_results_parameters_hidden}.

\begin{table}
    \centering
    \begin{tabular}{|c|c|c|c|c|c|c||c|}
    \hline
        & $\bm{N_0}$ $[\textrm{GeV} \cdot \textrm{s}]^{-1}$ & $\bm{\Gamma}_{\textrm{inj}}$ & $\bm{\tau}_{\textrm{d}}$ [yr] & $\bm{\alpha}_{\textrm{d}}$ & $\bm{t}_{\textrm{age}}$ [yr] & $\bm{d}$ [pc] & $\bm{E}_{\textrm{tot}}$ [erg] \\
        \hline 
        \hline
        \textbf{1 hidden} & $2.14 \cdot 10^{39}$ & $2.25$ & $1.13 \cdot 10^{5}$ & $2.40$ & $1.54 \cdot 10^{5}$ & $658.21$ & $2.45 \cdot 10^{49}$ \\
        \hline
        \textbf{4+1 hidden} & $6.10 \cdot 10^{39}$ & $2.05$ & $4.97 \cdot 10^3$ & $2.45$ & $4.97 \cdot 10^5$ & $1.19 \cdot 10^3$ & $1.94 \cdot 10^{49}$ \\
        \hline
    \end{tabular}
    \caption{\small{The table reports the MAP parameters resulting from the fit to the all-lepton flux. The \textbf{1 hidden} scenario identifies the case where only an unknown object is considered, while \textbf{4+1 hidden} fits an unknown SNR on top of the observed SNRs. The total energy injected by each source in the form of $e^{\pm}$ is also computed, based on the normalization.}}
    \label{tab:SNR_fit_results_parameters_hidden}
\end{table}

As a result of this analysis, we find that a hidden old remnant of $\sim 10^5$ yr is actually needed to reproduce correctly the data, and the best-fit distance is expected to be in the range ($600$ -- $1200$) pc. This range of distances is far from the one quoted in \cite{Recchia:2018jun}, where a very close source ($d = 100$ pc) is invoked to match the observed all-lepton data. The discrepancy is mainly due to the propagation model: we checked and found that, in accordance with \cite{Recchia:2018jun}, such a close source would correctly reproduce the data only if a diffusion coefficient with a Kolmogorov-like rigidity scaling ($\delta = 0.33$) and a smaller normalization were assumed. However, we exclude these parameters as they are not compatible with the observables we considered to calibrate our propagation setup ($p$, nuclei, B/C). Furthermore, we remark that our reference transport scenario with $\delta=0.45$ is consistent with the MCMC analysis carried out in \cite{Yuan:2017ozr}.

Given the required age, such a remnant would most likely be in its final \textit{radiative} phase and may be not clearly detectable (while this would be unrealistic for the much smaller distance found in \cite{Recchia:2018jun}). The SNR catalogue \cite{2012AdSpR..49.1313F} reports a possible candidate that we find particularly interesting, the Monogem Ring, which is categorized as \textit{uncertain SuperNova Remnant}. However, this source is too close ($d < 300$ pc) to the Earth and its propagated spectrum does not seem to be compatible with the high-energy ($E > 10$ TeV) all-lepton data, according to our propagation scenario. 

As a final discussion point on the electron spectrum, we point out that, as mentioned above, alternative explanations of the features we have analyzed so far have been recently put forward in the literature. In particular, in \cite{PhysRevLett.121.251106} an undiscovered pulsar is invoked to account for the all-lepton data. Another physical picture that requires a change in the propagation paradigm is found in \cite{Lipari:2018usj,Lipari:2019abu}. In that scenario the positron flux is entirely of secondary origin and the spectral break in the all-lepton spectrum is generated by energy-loss effects, possibly motivated by a much smaller ($\simeq 0.7 - 1.3$ Myr) residence time of the charged cosmic rays in the Galaxy compared to conventional scenarios; moreover, the break at $1$ TeV in the lepton spectrum would correspond to the energy at which the loss time becomes comparable with the diffusion timescale. We remark that a coherent picture that includes all the available hadronic and leptonic channels, together with gamma-ray and radio data, based on this idea has not been provided yet. However, it is an intriguing possibility that can be further tested with future, more accurate data, and a better characterization of the features discussed so far. In particular, we notice again that the presence of a spectral hardening at $\simeq 40$ GeV in the electron spectrum can be  interpreted, in our scenario, as the breakdown of the assumption of a continuous source term, and the signature of local sources that start to dominate the flux; on the other hand, this feature does not have a simple explanation in the alternative scenario based on purely secondary origin.

\section{Discussion and conclusions}

In this paper we provided a comprehensive discussion about the origin of the most relevant features observed in the positron, electron and all-lepton data recently released by the AMS-02, CALET, and H.E.S.S. Collaborations.

We first identified a CR transport scenario that very well reproduces B/C data published by AMS-02, and the proton, He, C and O data measured by AMS-02 and Voyager.

With this propagation setup at hand, we considered the positron data, that show a remarkable excess with respect to the secondary flux expected from the conventional proton-proton spallation process, and studied the expected contribution from individual pulsar wind nebulae (PWNe). Starting from a careful study of the analytical solution of the diffusion-loss equation from individual sources, we characterized the contribution due to a large number of old PWNe as a large scale extra-component which is often neglected in the related literature.
Then we focused on the prominent peak and drop-off in the positron spectrum recently found by AMS-02 around 300 GeV.
After pointing out that this feature is not compatible with alternative scenarios in which the largest part of the positron population is originated by CR nuclei scattering onto the ISM gas, we described it in terms of the emission from a young PWN under different conditions. 

We emphasize that we followed a different approach with respect to many other works on the same subject (see for instance \cite{DiMauro:2014iia} and \cite{Joshi:2017ogv}). 
In particular, given the poor knowledge of the emission mechanism details, we chose not to rely on a specific model but only on purely observational information, and performed a fit of the injection parameters for the extremal assumptions of burst-like and continuous injection, and for different injected spectral shapes.

We found that a hard acceleration spectrum and a spectral break or a cutoff at few hundred GeV are required to match the data, which is consistent with recent theoretical modeling of the typical acceleration mechanisms at the termination shock of PWNe.  

Finally, we turned our attention to the all-lepton spectrum and tried to reproduce its shape accounting for the contribution of known and possibly hidden SNRs. 

We pointed out that the contribution of local SNRs takes over the softer large-scale component at $\simeq 40$ GeV. 
We found however that the contribution of known nearby SNRs cannot reproduce the TeV feature recently identified by the H.E.S.S. Collaboration. This claim holds considering only statistical uncertainties, thus under the assumption that the systematics estimated by H.E.S.S. give rise to a rigid shift of the experimental points. Then, building on previous results from \cite{Recchia:2018jun}, we found that if a relatively near, old remnant is included in the calculation, with declining luminosity and with age $\sim 10^5$ yr and distance in the $600 - 1200$ pc range, the data are nicely reproduced within the propagation setup described in the first part, consistently with all the hadronic and leptonic channels under consideration. 

\acknowledgments

We are grateful to E. Amato, C. Evoli, S. Gabici, P. Lipari, P.S. Marrocchesi and S. Recchia for very inspiring discussions.

The work of DG has received financial support through the Postdoctoral Junior Leader Fellowship Programme from la Caixa Banking Foundation (grant n.~LCF/BQ/LI18/11630014).
DG was also supported by the Spanish Agencia Estatal de Investigaci\'{o}n through the grants PGC2018-095161-B-I00, IFT Centro de Excelencia Severo Ochoa SEV-2016-0597, and Red Consolider MultiDark FPA2017-90566-REDC.

OF warmly thanks the Instituto de F\'{i}sica Te\'{o}rica UAM-CSIC in Madrid, and in particular the DAMASCO group, for hosting him during the period January - June 2019 and La Caixa Banking Foundation (grant n.~LCF/BQ/LI18/11630014) for partial financial support. 

D. Grasso was supported by the grant ASI/INAF No. 2017-14-H.O.

We acknowledge the use of the IFT Hydra cluster for the development of part of this work.

\clearpage
\appendix
\section{Single-source solution to the transport equation}\label{app:solution_transport_equation}
In this appendix we review the analytical solutions to the transport equation in the different injection scenarios we consider throughout this work, as discussed in \cite{1995PhRvD..52.3265A,Ginzburg:1990sk}.


In the energy regime we are interested in (above $\sim 1$ GeV), we can rewrite the general form of Equation \eqref{eq:prop} in polar coordinates as follows:
\begin{equation}\label{eq:transport_equation_polar_coordinates_appendix}
    \frac{\partial f(E,t,r)}{\partial t} = \frac{D(E)}{r^2} \frac{\partial}{\partial r} r^2 \frac{\partial f}{\partial r} + \frac{\partial}{\partial E} (b(E) f) + Q,
\end{equation}
where $b(E)$ is the rate of energy-loss and $Q(E,t,\vec{r})$ the source term. In this energy regime the energy-loss is given by $b(E) = - b_0 E^2$, with $b_0 = 1.4 \cdot 10^{-16} \textrm{ GeV}^{-1} \cdot \textrm{s}^{-1}$.

Under the assumption that the emitting source is point-like, the Green-function approach to solve the equation gives the general solution: 
\begin{equation}\label{eq:solution_transport_equation_implicit}
    f(r,t,E) = \frac{Q(E_{\textrm{t}}) b(E_{\textrm{t}})}{\pi^{3/2} b(E) r^3_{\textrm{diff}}} \cdot e^{-\frac{r^2}{r^2_{\textrm{diff}}}},
\end{equation}
where we drop the dependence of the source term $Q$ on $t$ and $r$ for simplicity. $E_{\textrm{t}}$ refers to the energy at a time $(t - t_{\textrm{rel}})$ ago, that is $E_{\textrm{t}} = \frac{E}{1 - b_0 (t - t_{\textrm{rel}}) E}$. Therefore, the solution in Equation \eqref{eq:solution_transport_equation_implicit} becomes:
\begin{equation}\label{eq:solution_transport_equation_explicit}
    f(r,t,E) = \frac{Q(E_{\textrm{t}})}{\pi^{3/2} r^3_{\textrm{diff}}} \cdot \frac{1}{\left[ 1 - b_0 (t - t_{\textrm{rel}}) E \right]^2} \cdot e^{-\frac{r^2}{r^2_{\textrm{diff}}}},
\end{equation}
where $r^2_{\textrm{diff}} (E_{\textrm{t}},E) \equiv - 4 \int_{E_{\textrm{t}}}^{E} \frac{D(E')}{b(E')} dE'$ is the diffusive distance travelled by a particle loosing its energy from $E_{\textrm{t}}$ to $E$. This solution is still general, in that it does not contain any information about the injection term, that can be written $Q(t,r,E) = S(E) L(t) \delta(r)$.
\vspace{0.3cm}

\hspace{-0.9cm}\textbf{\textit{Decaying-luminosity injection}}. When no further information is provided on the luminosity timescale, the decaying-luminosity function is in the general form $L(t) = \frac{ L_0 }{ \left(  1 + \frac{t}{\tau_{\text{d}}}  \right)^{\alpha_{\text{d}}}}$, with $\alpha_{\textrm{d}}$ and $\tau_{\textrm{d}}$ parameters characteristic of the emission mechanism. Integrating over time the expression \eqref{eq:solution_transport_equation_explicit}, we obtain:
\begin{equation}\label{eq:solution_tr_equation_general_form}
    f(r,t_{\textrm{age}},E) = \int_{t_{\textrm{rel}}}^{t_{\textrm{age}}} dt' \frac{S(E_{t'}) L(t') }{\pi^{3/2} r_{\textrm{diff}}^3(E,E_{t'})} \cdot \frac{1}{\left[ 1 - b_0 (t_{\textrm{age}} - t') E \right]^2} \cdot e^{-\frac{r^2}{r^2_{\textrm{diff}}}} \, ,
\end{equation}
where $t_{\textrm{rel}}$ is the release time of the particles.

\vspace{0.3cm}

\hspace{-0.9cm}\textbf{\textit{Constant-luminosity injection}}. When the luminosity timescale $\tau_0$ is much larger than the age of the source, we can approximate $L(t) \rightarrow L_0 dt$ and the solution takes the form:
\begin{equation}\label{eq:solution_constant_luminosity}
    f(r,t_{\textrm{age}},E) = \frac{L_0 S(E)}{4 \pi D(E) r} \cdot \textrm{erfc} \left( \frac{r}{\sqrt{4 D(E) (t_{\textrm{age}} - t_{\textrm{rel}})}} \right),
\end{equation}
with $\textrm{erfc} (x) = \frac{2}{\sqrt{\pi}} \int_x^{\infty} e^{-t^2} dt$ the complementary error-function.
\vspace{0.3cm}

\hspace{-0.9cm}\textbf{\textit{Burst-like injection}}. When $\tau_0$ is much smaller than the age of the source, the luminosity function tends to $L(t) \rightarrow L_0 \delta(t - t_{\textrm{rel}}) dt$, and the solution \eqref{eq:solution_tr_equation_general_form} basically takes the form of the integrand function:
\begin{equation}\label{eq:solution_burst_injection}
    f(r,t_{\textrm{age}},E) = \frac{S(  E_{t_{\textrm{age}}}  )}{\pi^{3/2} r_{\textrm{diff}}^3(E,E_{t_{\textrm{age}}})} \cdot \frac{1}{\left[ 1 - b_0 (t_{\textrm{age}} - t_{\textrm{rel}}) E \right]^2} \cdot e^{-\frac{r^2}{r^2_{\textrm{diff}}}}.
\end{equation}


The decaying-luminosity and burst-like solutions are valid as long as the condition $E < \frac{1}{b_0 ( t_{\textrm{age}} - t_{\textrm{rel}} )}$ holds.

Throughout this work we will implement the following source features:
\begin{itemize}
    \item[-] power-law with cutoff: $S(E) = S_0 \left( \frac{E}{E_0} \right)^{\Gamma_{\textrm{inj}}} \cdot e^{-\frac{E}{E_{\textrm{cut}}}}$
    \item[-] broken power-law: $S(E) = S_0 \left( \frac{E}{E_0} \right)^{\Gamma_{\textrm{inj}}} \cdot \left(  1 + \left( \frac{E}{E_{\textrm{break}}}  \right)^{|\Delta \Gamma_{\textrm{inj}}| \cdot \textrm{s}}   \right)^{\textrm{sign}(\Delta \Gamma_{\textrm{inj}})/\textrm{s} }$,
\end{itemize}
where $\Delta \Gamma_{\textrm{inj}}$ is the change in the injection index and $s$ a parameter that rules the sharpness of the change in the slope. 


\section{Estimation of the release time from PWNe}\label{app:release_time_from_PWN}

Since the release of the PAMELA data on the positron fraction, several phenomenological scenarios invoked a relevant delay between pulsar formation and the release of the electron+positron pairs in the ISM (see for instance \cite{2009APh....32..140G}). The physical picture behind this time delay, extensively discussed for example in \cite{2011ASSP...21..624B}, is the following. A typical pulsar forms in a core collapse supernova event with a natal kick velocity of 
$\simeq 400$ km/s or larger; this relevant proper motion drives the compact object far from the place of its formation, across the supernova remnant and then across the shocked ejected material. After the escape from the remnant, as a consequence of the impact
of the relativistic PWN wind onto the ISM, a bow shock forms. Such structure can hardly confine the electron+positron pairs accelerated within the PWN: the particles can hence escape from the PWN and contribute to the diffuse sea of cosmic radiation.

In this work, guided by this physical picture, we estimate the release time by computing the time needed by a pulsar with a typical kick velocity to escape a typical SN Ia remnant. The time evolution of the SNR shock radius is computed following the prescriptions summarized in \cite{Gaggero:2017abc}. In particular, the ejecta-dominated phase is described by the self-similar solutions provided by \cite{Chevalier1982ApJ}, and the subsequent Sedov phase is modeled adopting the thin-shell approximation \cite{Ostriker1988}, based on the assumption that the mass is mostly concentrated within a shell of negligible thickness at the forward shock. 
Given these assumptions on the SNRs, and within a wide range of pulsar kicks, spanning from $100$ to $1000$ km/s, we obtain release times in the interval [$10^4$ -- $5 \cdot 10^5$] yr. For the purpose of this work, we consider an intermediate reference value $t_{\textrm{rel}} = 6.4 \cdot 10^4$ yr, that corresponds to a pulsar with kick $v_{\textrm{pulsar}} = 400$ km/s.

\section{Notes on the pulsars from ATNF Catalogue}\label{app:appendix_pulsars_ATNF}

The position of the peak in the positron flux ($\sim 250$ GeV) requires sources that are as old as $\sim 10^6$ yr, based on $E_{\textrm{peak}} = 1/(b_0 \cdot (t_{\textrm{ch}} - t_{\textrm{rel}}))$. A particle diffusing in the Galaxy for this time interval is coming from a distance $\sqrt{4 \cdot D(E_{\textrm{peak}}) \cdot (t_{\textrm{ch}} - t_{\textrm{rel}})} \simeq 1.3$ kpc. 

\begin{figure}
    \centering
    \includegraphics[width=0.65\linewidth]{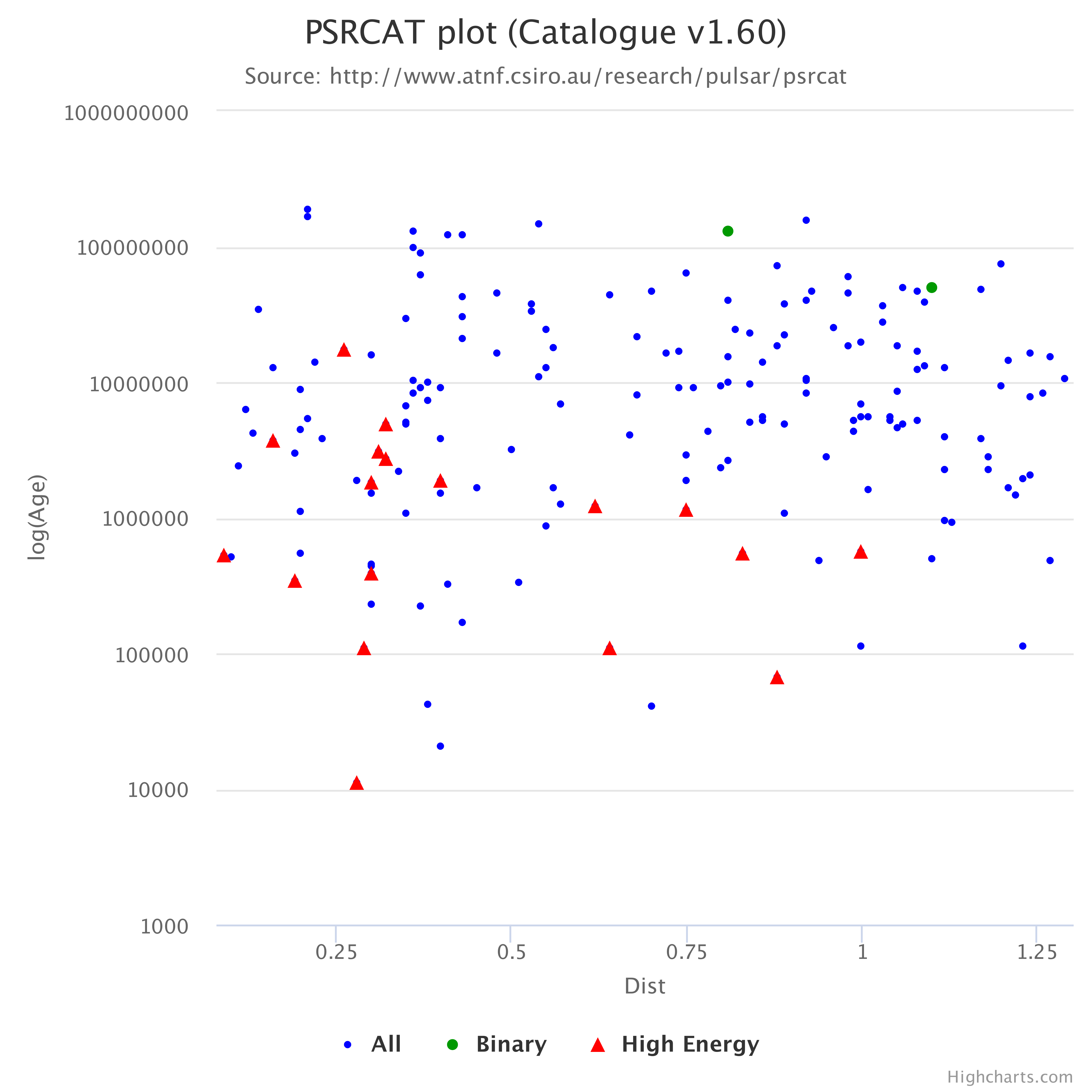}
    \caption{\small{The figure is a ($d, \log_{10}(t_{\textrm{age}})$) scatter plot of all the pulsars in the ATNF Catalogue within $1.3$ kpc and younger than $2 \cdot 10^{8}$ yr. Marked with red triangles there are high-energy pulsars, that have an emission at frequency higher than infrared. As they are distributed quite uniformly, we will consider them as a good sample of pulsar population}.}
    \label{fig:pulsar_population_ATNF}
\end{figure}

In Figure \ref{fig:pulsar_population_ATNF} we report all the pulsars listed in the ATNF Catalogue that are found within this distance and younger than $2 \cdot 10^8$ yr. We make them inject leptons with a hard spectrum ($\Gamma_{\textrm{inj}} = 1.5$) up to an energy $E_{\textrm{cut}} = 300$ GeV, where an exponential cutoff $e^{-\frac{E}{E_{\textrm{cut}}}}$ is implemented. This is consistent with \cite{Amato:2013fua}, where it is argued that pulsar emission requires an injection break due to a change in the accelerating site around the compact object: leptons up to $200$ -- $400$ GeV are accelerated within the nebula by mechanisms that are not fully understood (\textit{e.g.} magnetic reconnection), with a hard injection $\Gamma_{\textrm{inj}} < 2$, while more energetic leptons are accelerated at the termination shock, thus with a softer spectrum $\Gamma_{\textrm{inj}} > 2$ characteristic of the DSA. It is not clear whether the second population can be considered subdominant, thus justifying a cutoff instead of a break. However, this does not affect much the energy budget injected by the source. After the injection, we make them propagate through the Galaxy via the transport-equation \eqref{eq:transport_equation_polar_coordinates_appendix}.

For the release time of the leptons, we consider the value $t_{\textrm{rel}} = 6.4 \cdot 10^4$ yr, corresponding to a pulsar with birth speed $v_{\textrm{pulsar}} = 400$ km/s, as described in Appendix \ref{app:release_time_from_PWN}. We verified that the extreme values discussed there do not change qualitatively the results.
We observe that a different release time from the PWN effectively mimics the effect of a surrounding confinement region, like the ones observed by HAWC \cite{2017Sci...358..911A}, although we are not taking into account the possible reshaping of the spectral index introduced by losses inside those regions.

\begin{figure}
    \begin{subfigure}{.5\linewidth}
        \centering
        \includegraphics[width=1.\linewidth]{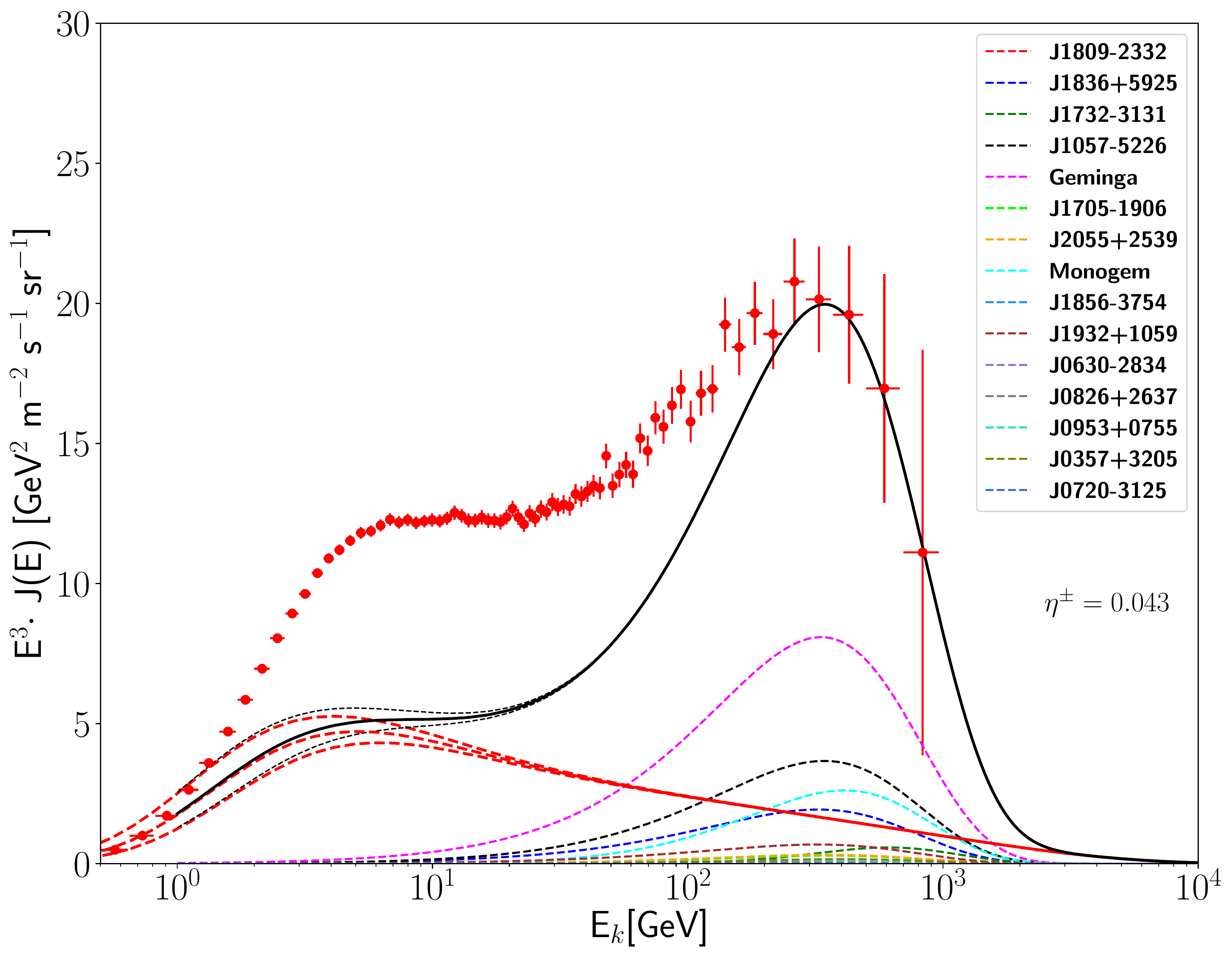}
        \caption{}
        \label{subfig:pulsar_contribution_plus_secondary_constant_L_HE}
    \end{subfigure}
    \begin{subfigure}{.5\linewidth}
        \centering
        \includegraphics[width=1.\linewidth]{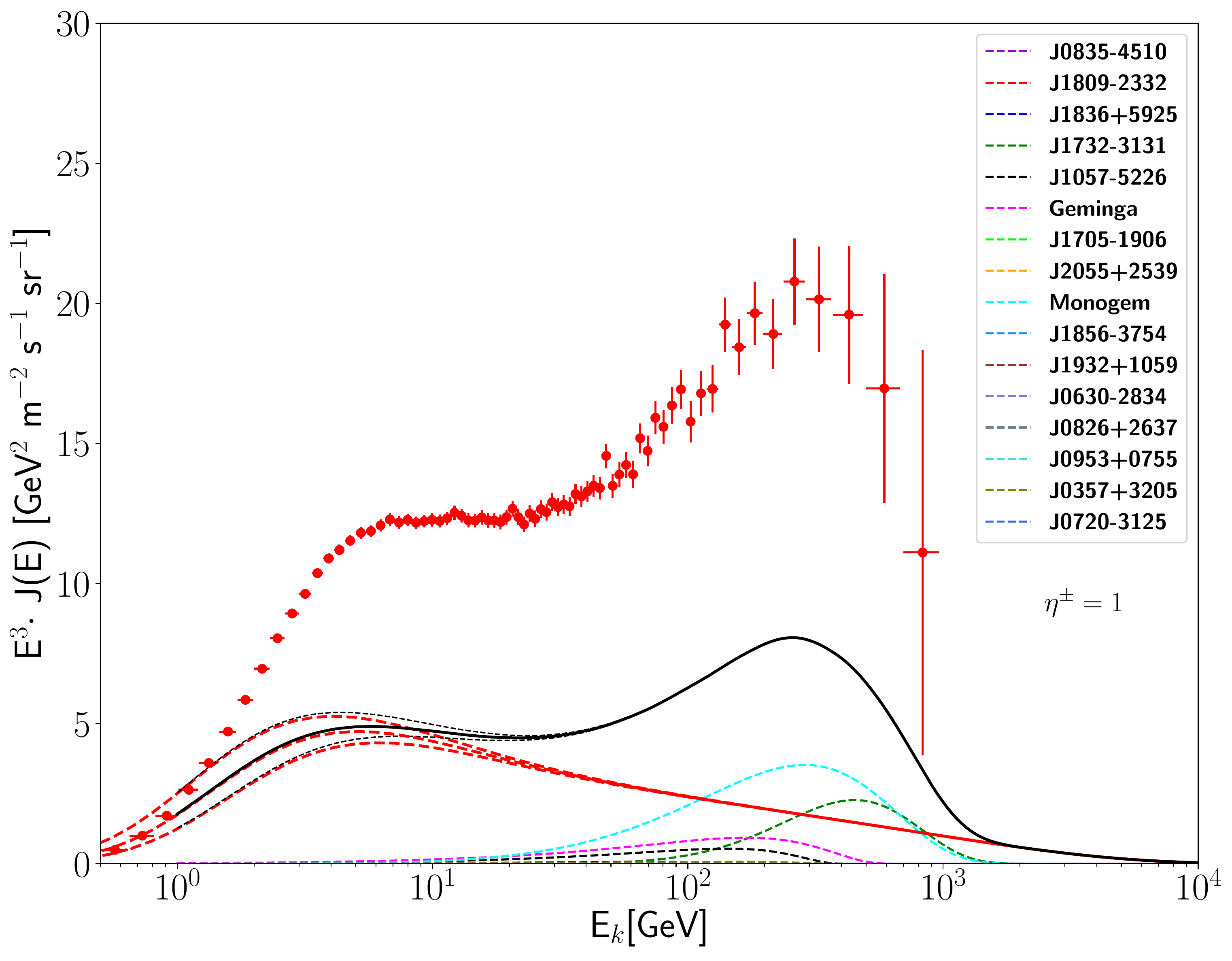}
        \caption{}
        \label{subfig:pulsar_contribution_plus_secondary_burst_HE}
    \end{subfigure}
\caption{\small{We plot here all the high-energy pulsars within $1.3$ kpc and younger than $2 \cdot 10^{8}$ yr found in Figure \ref{fig:pulsar_population_ATNF}. (a) Sources are propagated from a constant-luminosity injection: the high-energy data are reproduced with a conversion efficiency of $\eta^{\pm} = 0.043$. (b) Sources are propagated from a burst-like injection: the high-energy data cannot be matched, due to the insufficient nominal injected energy. Notice the inverted hierarchy of the dominant sources.}}
    \label{fig:pulsars_from_ATNF_HE}
\end{figure}

With this emission paradigm, we plot all the sources that in Figure \ref{fig:pulsar_population_ATNF} are marked as \textit{high-energy pulsars}. This denomination is due to the emission frequency, but we consider them because they uniformly span the scatter plot and thus constitute a good sample. The result is shown in Figure \ref{fig:pulsars_from_ATNF_HE}, where the constant-luminosity (\ref{subfig:pulsar_contribution_plus_secondary_constant_L_HE}) and the burst-like (\ref{subfig:pulsar_contribution_plus_secondary_burst_HE}) solutions to \eqref{eq:transport_equation_polar_coordinates_appendix} are compared: only the constant luminosity injection can reproduce the positron data. This can be due to the total amount of injected energy, that we estimated trivially as $E_{\textrm{tot}} = \lvert \dot{E}_{\textrm{loss}} \rvert \cdot t_{\textrm{ch}}$. 

As it can be easily understood, this is a lower bound (LB), since it is based on the current measurements of $\Omega$ and $\dot{\Omega}$ in $\dot{E}_{\textrm{loss}}= - I\Omega \dot{\Omega}$. In fact the rotational frequency $\Omega$ is currently smaller than at the beginning of its life, as well as its variation $\dot{\Omega}$. We can do an attempt to improve the estimation for $E_{\textrm{tot}}$ by implementing the magnetic dipole (MD) radiation model, as follows:
\begin{equation}\label{eq:magnetic_dipole_total_energy_formula}
    E_{\textrm{tot,MD}} = \int_{t_{\textrm{rel}}}^{t_{\textrm{age}}} \lvert \dot{E}_{\textrm{MD}} \rvert \, dt,
\end{equation}
where $\dot{E}_{\textrm{MD}} = - \frac{B^2 R^6 \Omega^4}{6 c^3}$, $\Omega(t) = \frac{\Omega_0}{\sqrt{1 + \frac{t}{\tau_{0,\textrm{MD}}}}}$.

Carrying out the integral, we obtain:
\begin{equation}\label{eq:magnetic_dipole_total_energy_integrated}
    E_{\textrm{tot,MD}} = \tau_{0,\textrm{MD}} \frac{B^2 R^6 \Omega_{0}^{4}}{6 c^3} \cdot \left(    \frac{1}{1 + \frac{t_{\textrm{rel}}}{\tau_{0,\textrm{MD}}}} -  \frac{1}{1 + \frac{t_{\textrm{age}}}{\tau_{0,\textrm{MD}}}} \right).
\end{equation}

With the ATNF parameters, we find $E_{\textrm{tot,MD}} < E_{\textrm{tot,LB}}$, which is a hint that the emission mechanism requires some modification. Different values of $t_{\textrm{rel}}$ do not affect this conclusion.

Regardless, there are two model-independent aspects that we observe:
\begin{enumerate}
    \item There is a very different conversion efficiency for the two injection scenarios, compatibly with what is discussed in Section \ref{subsec:bayesian_fit_positrons}.
    \item Among the dominant sources, the hierarchy is inverted between Monogem and Geminga: this is expected as well, if one considers the interplay among the nominal parameters of the two pulsars. In fact, when particles are injected instantaneously (burst-like), the younger source dominates over the older one, as particles had less time to loose energy. On the other hand, for the constant-luminosity case, the sources are still emitting, therefore the discriminating parameter here is $E_{\textrm{tot}}$.
\end{enumerate}
In both cases, one source dominates by a factor of $\sim 2$, and this favors our parametrization of the fits in Section \ref{subsec:bayesian_fit_positrons} with one prominent source. 




\clearpage
\bibliographystyle{unsrt}
\bibliography{Paper_features_in_CR_leptons.bib}

\end{document}